\documentclass{article}
\usepackage[a4paper, total={6.5in, 8in}]{geometry}

\usepackage{hyperref}
\usepackage{url}

\usepackage{microtype}
\usepackage{graphicx}
\usepackage{booktabs} % for professional tables

\usepackage{hyperref}
\usepackage[textsize=tiny]{todonotes}

\usepackage{amsthm}
\usepackage{amsmath,amssymb,amsfonts}
\usepackage{algorithmic}
\usepackage{algorithm2e}
\usepackage{graphicx}
\usepackage[capitalize,noabbrev]{cleveref}

\usepackage{hyperref}
\hypersetup{
    colorlinks=true,   
 }

\usepackage{textcomp}
\usepackage[T1]{fontenc}
\usepackage{color}
\usepackage{xurl}
\usepackage{bm}
\usepackage[normalem]{ulem}
\usepackage{xcolor}
\usepackage{xspace}
\usepackage{wrapfig}
\usepackage{booktabs, multirow}

\usepackage{enumitem}
\usepackage{subcaption}
\usepackage{caption}
\usepackage{float}

\usepackage{thmtools}
\usepackage{thm-restate}
\usepackage{Definitions}
\usepackage{amsmath,amsfonts,bm}

\def\eqref#1{equation~\ref{#1}}
\def\1{\bm{1}}

\def\rb{{\textnormal{b}}}

\def\vb{{\bm{b}}}

\DeclareMathAlphabet{\mathsfit}{\encodingdefault}{\sfdefault}{m}{sl}
\SetMathAlphabet{\mathsfit}{bold}{\encodingdefault}{\sfdefault}{bx}{n}

\def\gA{{\mathcal{A}}}

\def\gG{{\mathcal{G}}}

\def\gS{{\mathcal{S}}}

\def\sN{{\mathbb{N}}}

\newcommand{\wM}{\widetilde{M}}
\newcommand{\wQ}{\widetilde{Q}}

\newcommand{\wH}{\widetilde{H}}

\let\ab\allowbreak

\newcommand\bbR{\ensuremath{\mathbb{R}}} % Real numbers
\newcommand\bbE{\ensuremath{\mathbb{E}}} % Expectation
\newcommand\bbP{\ensuremath{\mathbb{P}}} % Probability
\newcommand{\ep}{\epsilon} % epsilon
\newcommand{\xpt}[2][]{\bbE_{#1}\left[{#2}\right]}
\newcommand{\prob}[1]{\bbP\left({#1}\right)}

\DeclarePairedDelimiterX{\infdivx}[2]{(}{)}{%
  #1\;\delimsize\|\;#2%
}
\newcounter{relctr}[section] %% <- counter for relations

\newcommand\labelrel[2]{%
  \begingroup
    \refstepcounter{relctr}%
    \stackrel{\textnormal{(\roman{relctr})}}{\mathstrut{#1}}%
    \originallabel{#2}%
  \endgroup
}
\AtBeginDocument{\let\originallabel\label}

\newtheorem{assump}{Assumption}

\newboolean{showcomments}
\setboolean{showcomments}{true}
\allowdisplaybreaks

\usepackage{authblk}

\usepackage{thmtools}
\usepackage{thm-restate}

\crefname{condition}{Condition}{Condition}

\allowdisplaybreaks

\author{}

\title{Scalable spectral representations for multi-agent \\
reinforcement learning in network MDPs}
\author{Zhaolin Ren$^{*,1}$, Runyu (Cathy) Zhang$^{*,1}$, Bo Dai$^{2}$ and  Na Li$^{1}$
    \thanks{$^*$Equal contributions; $^{1}$Z. Ren, R. Zhang, and  N. Li are with Harvard University, Email: {zhaolinren@g.harvard.edu, runyuzhang@fas.harvard.edu, nali@seas.harvard.edu}; $^2$B. Dai is with Google Brain and Georgia Tech, Email: {bodai@\{google.com, cc.gatech.edu\}}.}
    \thanks{
    Z. Ren, R. Zhang, and N. Li are funded by NIH R01LM014465, NSF AI institute: 2112085, and NSF ASCENT: 2328241}
}

\makeatletter
\def\thanks#1{\protected@xdef\@thanks{\@thanks
        \protect\footnotetext{#1}}}
\makeatother

\begin{document}

\maketitle

\begin{abstract}
Network Markov Decision Processes (MDPs), a popular model for multi-agent control, pose a significant challenge to efficient learning due to the exponential growth of the global state-action space with the number of agents. 
In this work, utilizing the exponential decay property of network dynamics, we first derive scalable spectral local representations for network MDPs, 
% which have enabled efficient learning for single-agent MDPs, 
which induces a network linear subspace for the local $Q$-function of each agent. 
Building on these local spectral representations, we design a scalable algorithmic framework for continuous state-action network MDPs, and provide end-to-end guarantees for the convergence of our algorithm. Empirically, we validate the effectiveness of our scalable representation-based approach on two benchmark problems, and demonstrate the advantages of our approach over generic function approximation approaches to representing the local $Q$-functions.

% \lina{We can either make a short abstract by not talking too much of the background/literature or if we talk, we should talk about the work which did function approximation--using a NN to approximate Q for continuous state/action.} \zhaolin{I made it shorter.}
\end{abstract}

%%%%%%%%%%%%%%%%%%%%%%%%%%%%%%%%%%%%%%%%%%%%%%%%%%%%%%%%%%
% \vspace{-1em}
\section{Introduction}\label{sec:intro}
% \vspace{-1em}
%%%%%%%%%%%%%%%%%%%%%%%%%%%%%%%%%%%%%%%%%%%%%%%%%%%%%%%%%%
% Paper outline:
% \begin{enumerate}
%     \item intro
    
%     \item problem setup and prelim; mention exponential decay property, previous works study how this can be used to enable scalable control by using truncated local Q-functions. However, pose the question: what about continuous state-action space? Naively using deep NN may lead to inefficiencies/difficult to tune, etc.
%     \item representing truncated local Q-function in continuous state-action space.
%     \begin{enumerate}
%         \item From linear MDP, we know that if P is linearly factorizable, then Q admits linear representation.
%         \item However, in network case, this linear representation may scale exponentially with n. 
%         \item So we leverage instead truncating the kappa-hop dynamics, which combined with the exponential decay property, allows us to approximate the $Q_i$ function using truncated local features. Show result in exact case.
%         \item Then, talk about result when P is not exactly factorizable. In this case, when local dynamics are Gaussian, then the localized P can be approximated using finite-dim spectral features, e.g. random features. Defer details to appendix. State the result.
%     \end{enumerate}
%     \item show policy evaluation error
%     \item show policy optimization error (in the multi-agent case, converge to neighborhood of Nash equilibrium)?
%     \item Numerical results
%     \item Conclusion
% \end{enumerate}

Multi-agent network systems have found applications in various societal infrastructures, such as power systems, traffic networks, and smart cities~\cite{mcarthur2007multi,burmeister1997application,roscia2013smart}. One particularly important class of such problems is the cooperative multi-agent network MDP setting, where agents are embedded in a graph, and each agent has its own local state~\cite{pmlr-v120-qu20a}. In network MDPs, the local state transition probabilities and rewards only depend on the states and actions of the agent's \emph{direct neighbors} in the graph. Such a property has been observed in a great variety of cooperative network control problems, ranging from thermal control of multizone buildings~\cite{zhang2016decentralized}, wireless access control~\cite{zocca2019temporal} to phase synchronization in electrical grids~\cite{blaabjerg2006overview}, where agents typically only need to act and learn based on information within a local neighborhood due to constraints on the information and communication infrastructure. 
However, despite many efforts (c.f. \cite{qu_scalable_2021,lin_multi-agent_2021,zhang2023global,abdallah_multiagent_2007,du_scalable_2022,ma_efficient_2024}), efficiently finding effective local policies for networks remains an open challenge. 

% \lina{``In these networks, one agent .... `` setting up the network problem: local interaction} \lina{Due to the information and communication infrastructure, agents need to make their own decision based on local information.} \lina{However, despite many efforts \cite{references in network control, distributed control}, finding efficient local policies for networks remains an open challenge given the complexity of the systems.} 

Reinforcement Learning (RL)~\cite{sutton2018reinforcement} has emerged as a promising tool for addressing the complex dynamics of these systems \cite{chen_communication-efficient_2024,nezamoddini_survey_2022,yan_multi-agent_2020}. 
There are several pioneering works on designing scalable RL algorithms for network systems \cite{qu_scalable_2021,lin_multi-agent_2021,zhang2023global}. To facilitate scalable control in network control, in \cite{qu_scalable_2021}, the authors introduced a key insight, referred to as the \emph{exponential decay property} of the Q-function. This property suggests that each agent's local Q-function can be well-approximated using only information from its $\kappa$-hop neighborhood. We note that a similar property has also been proposed in \cite{gu_mean-field_2022} which focuses on reinforcement learning in the mean field multi-agent setting. Leveraging this property, the proposed algorithm concentrates on learning truncated Q-functions and then applying either policy gradient \cite{qu_scalable_2021,lin_multi-agent_2021} or policy iteration \cite{zhang2023global}. However, although these methods are scalable with respect to the network size, they are limited to the tabular setting, where each agent must store a local $Q$-table that scales with the state and action spaces of its neighborhood, making it inefficient for large state and action spaces. 
In fact, due to the inherent complexity in network MDPs, \ie, \emph{network size} is large and the \emph{state and action spaces} of each agent are large or continuous, designing efficient and scalable RL algorithms for such systems remains a long-standing challenge.
% \textcolor{blue}{A similar exponential decay property is also proposed in \cite{gu_mean-field_2022} which focuses on reinforcement learning in the mean field multi-agent setting.} 

There have been several works aimed at addressing scalability in the context of large state and action spaces. A common approach is to use function approximation to find an efficient representation of the Q-function. For instance, \cite{stankovic_multi-agent_2016} explores linear function approximation to solve network RL problems. However, their setting is simpler than the network MDP considered here, as they assume fully decoupled agent dynamics, whereas we allow an agent's dynamics to depend on the states of neighboring agents. In the broader context of multi-agent learning, linear function approximation has also been widely studied \cite{zhang_fully_2018,dubey_provably_2021}. However, these works differ from ours: \cite{zhang_fully_2018} focuses on the stochastic game setting, where agents share a common global state, while \cite{dubey_provably_2021} examines the parallel MDP setting. Additionally, outside the RL domain, there are works on network representation learning \cite{dong_heterogeneous_2020,li_network_2020,zhang_network_2020}. However, it remains unclear whether these techniques can be applied to control and RL in network systems, which presents an interesting open question for future research.

Finding a suitable representation for the Q-function is not a unique problem in network RL. It is also a central challenge in classical centralized or single-agent RL. Recently, there have been several papers that discuss function approximation in the centralized RL regime \cite{jin_provably_2020,du_bilinear_2021,uehara_representation_2022,grunewalder_modelling_nodate}. For example, \cite{jin_provably_2020} focuses on the linear MDP setting, where the transition kernel of the MDP can be represented as a linear combination of low-rank features. 
% \runyu{some terminology questions, can we abbreviate `linear combination of features' with `linear features'?}\zhaolin{I think linear combo is OK. Linear features might confuse some people.}
They demonstrate that in this setting, scalable RL can be achieved, with sample complexity depending on the size of the feature space rather than the size of the state and action spaces. Linear MDPs also have broad applications. Notably, \cite{ren_free_2022} explore the connection between stochastic nonlinear dynamical systems and linear MDPs, showing that under certain noise assumptions, stochastic nonlinear dynamics can be well-approximated by linear MDPs through an approach called \emph{spectral dynamic embedding}. Building on this, \cite{ren_stochastic_2023} developed RL algorithms specifically for linear MDPs.

Given the existing literature, the following question remains open:
% \bo{the question is too long.....}
\begin{center}
% \emph{How do we identify an appropriate representation for continuous state-action network MDPs and leverage it to achieve scalability with respect to both the network size and the size of the state and action spaces?}    
% \emph{Can we bypass the curse of dimensionality for continuous state-action network MDPs?}
% \runyu{I actually prefer the original question because it is more precise. I made another attempt, feel free to use it or not}
\vspace{-5pt}
\emph{Can we identify an appropriate representation for network MDPs and leverage it for scalability in both the size of the network and state-action space?}
\end{center}

\paragraph{Our contribution} Building on the existing literature, this paper addresses the critical gap by proposing a spectral dynamic embedding-based representation and developing a multi-agent RL algorithm for network systems that \emph{scales efficiently with both network size and the complexity of state and action spaces}, while also providing provable convergence guarantees.

Our approach integrates insights from both network RL and scalable centralized RL. Specifically, utilizing the exponential decay property and local nature of the transition dynamics, we show how we can approximate the local $Q_i$-value function linearly via \emph{network $\kappa$-local spectral features} that factorize the $\kappa$-hop transition dynamics. Leveraging this property, we develop a scalable sample-efficient method to learn local Q-functions in continuous network MDPs, followed by policy optimization based on the learned $Q$-functions.
% A key observation of our algorithm, built upon the \emph{exponentially decaying property} \cite{qu_scalable_2021}, is that each agent’s local Q-function can be well-approximated by the concatenation of its $\kappa$-hop neighbors’ linear features. 

% Moreover, we highlight the broad applicability of the \emph{network linear MDP} model by demonstrating how a \emph{spectral dynamics embedding} approach, similar to that in \cite{ren_free_2022}, can be used to approximate complex network dynamical systems as network linear MDPs. 
We provide rigorous sample complexity guarantees for our framework, and to the best of our knowledge, this is the first work to propose a provably efficient multi-agent RL algorithm for network systems that is scalable with respect to both network size and the size of the state and action spaces of individual agents. Finally, we validate our approach with numerical experiments on network thermal control and Kuramoto oscillator synchronization. In both cases, we find that our approach provides benefits over generic neural network function approximations, demonstrating the advantages of our spectral representation-based framework.
\vspace{-5pt}
\paragraph{Notations} For any vectors $v_1,\dots, v_n \in \bbR^d$, the notation $\otimes_{i=1}^n v_i \in \bbR^{nd}$ denotes their tensor product. The inner product of two tensor products is defined as follows. Consider another set of vectors $w_1,\dots,w_n \in \bbR^d$. Then, we denote $\inner{\otimes_{i=1}^n v_i}{\otimes_{i=1}^n w_i}  := \prod_{i=1}^n \inner{v_i}{w_i}.$ We also use the notation $[n]$ to denote the set $\{1,\dots,n\}$ for a positive integer $n$. In addition, when the context is clear, for notational convenience, we may drop the time indices and denote $(s(t),a(t), s(t+1))$ as $(s,a, s')$.

%%%%%%%%%%%%%%%%%%%%%%%%%%%%%%%%%%%%%%%%%%%%%%%%%%%%%%%%%%
\section{Problem Setup and Preliminaries}\label{sec:prelim}
%%%%%%%%%%%%%%%%%%%%%%%%%%%%%%%%%%%%%%%%%%%%%%%%%%%%%%%%%%

% \bo{preliminaries section is too long. Please reduce this section and leave only necessary brief introduction to network MDP. 
% \begin{itemize}
%     \item 1, Example of Kuramoto oscillator synchronization can be moved to appendix.  
%     \item 2, the discussion about exponential decay is also too long. You just need to specify what is exponential decay and why that is necessary. 
% \end{itemize}
% Please consider to shrink this section at most 1 page. 
% }
% \zhaolin{Thanks. Will shrink this part.}

% \lina{general comments: the writing needs revision, especially on the story. When we introduce the network mdp, we should immediately provide examples to explain; when we talk about the  "spectral assumptions/properties", we should have examples too, even if it is about refer to the TAC paper about the stochastic nonlinear dynamics, Then we also need to have a story line: saying that if the dynamics is this, we will show that $Q$ representation. For the $Q$ representation, we should make it clear that two questions: 1) global $Q$ to local $Q$; 2) spectral space.}

% \lina{because the deadline is in a week, I am not sure whether we will have enough time to make a solid draft. For ML conferences, I think writing the story carries too much weight...}

\paragraph{Network Markov Decision Process (MDP)} We consider the network MDP model, where there are $n$ agents associated with an underlying undirected graph \( \gG = (\mathcal{N}, \mathcal{E}) \), where \( \mathcal{N} = \{1, \ldots, n\} \) is the set of agents and \( \mathcal{E} \subseteq \mathcal{N} \times \mathcal{N} \) is the set of edges. Each agent \( i \) is associated with state \( s_i \in \gS_i \), \( a_i \in \gA_i \) where \( \mathcal{S}_i \subset \bbR^S \) and \( \gA_i \subset \bbR^A \) are bounded compact sets. At each time $t \in \sN$, the global state of the network is denoted as $s(t) = (s_1(t),\dots,s_n(t)) \in \mathcal{S} := \mathcal{S}_1 \times \dots \mathcal{S}_n$. Similarly, the global actuation of the network at each time $t$ is denoted as $a(t) = (a_1(t),\dots,a_n(t)) \in \mathcal{A} := \mathcal{A}_1 \times \dots \mathcal{A}_n$. We also introduce the following notations related to \emph{$\kappa$-hop neighborhoods}. Let \( N_i^\kappa \) denote the set of \( \kappa \)-hop neighborhood of node \( i \) and define \( N_{-i}^\kappa = \mathcal{N} \setminus N_i^\kappa \), i.e., the set of agents that are outside of \( i \)'th agent's \( \kappa \)-hop neighborhood. We write state \( s \) as \( (s_{N_i^\kappa}, s_{N_{-i}^\kappa}) \), i.e., the states of agents that are in the \( \kappa \)-hop neighborhood of \( i \) and outside of \( \kappa \)-hop neighborhood respectively. Similarly, we write \( a \) as \( (a_{N_i^\kappa}, a_{N_{-i}^\kappa}) \). When $\kappa = 1$, for simplicity we denote $N_i := N_i^1$.

We assume that the next state of each agent $i$ depends only on the current states and actions of its neighbors, so that the probability transition admits the following factorization
\begin{align*}
    \label{eq:P_dynamics_decomp}
    \prob{s(t+1) \mid s(t),a(t)} = \prod_{i=1}^n \prob{s_i(t+1) \mid s_{N_i}(t), a_{N_i}(t)},
\end{align*}
where $N_i$ indicates the neighbors of agent $i$, and $s_{N_i}(t)$ denotes the states of the neighbors of agent $i$ at time $t$. Further, each agent is associated with a stage reward function \( r_i(s_{N_i}, a_i) \) that depends on the local state and action, and the global stage reward is \( r(s, a) = \frac{1}{n} \sum_{i=1}^{n} r_i(s_{N_i}, a_i) \); for simplicity, in the rest of our paper, we will assume that $r_i$ depends only on $(s_i,a_i)$, but we note that our analysis carries with minimal changes when $r_i$ depends on $(s_{N_i},a_i)$. The objective is to find a (localized) policy tuple \( \pi = (\pi_1, \ldots, \pi_n) \), where each $\pi_i(\cdot \mid s) \equiv \pi_i(\cdot \mid s_{N_i^{\kappa_\pi}})$ depends only on a $\kappa_\pi$-hop neighborhood, such that the discounted global stage reward is maximized, starting from some initial state distribution \( \mu_0 \),
\begin{align*}
\max_{\pi} J(\pi) := \mathbb{E}_{s \sim \mu_0} \mathbb{E}_{a(t) \sim \pi(\cdot|s(t))} \left[ \sum_{t=0}^{\infty} \gamma^t r(s(t), a(t)) | s(0) = s \right].
\end{align*}
 
Next, we give the Kuramoto oscillator synchronization problem as an example of continuous state-action network MDPs. We defer another example, that of thermal control of multi-zone buildings, to Appendix \ref{appendix:more_network_mdp_examples}.

\begin{example}[Kuramoto oscillator synchronization]
The Kuramoto model \cite{acebron2005kuramoto,dorfler2012synchronization} is a well-known model of nonlinear coupled oscillators, and has been widely applied in various fields, ranging from synchronization of neurons in the brain~\cite{cumin2007generalising}, to synchronization of frequency of the alternating current (AC) generators or oscillators~\cite{filatrella2008analysis}. Concretely, we consider here a Kuramoto system with $n$ agents, with an underlying graph \( \gG = (\mathcal{N}, \mathcal{E}) \), where \( \mathcal{N} = \{1, \ldots, n\} \) is the set of agents and \( \mathcal{E} \subseteq \mathcal{N} \times \mathcal{N} \) is the set of edges. The state of each agent \( i \) is its phase \( \theta_i \in [-\pi,\pi] \), and the action of each agent is a scalar \( a_i \in \gA_i \subset \bbR\) in a bounded subset of $\bbR$. The dynamics of each agent is influenced only by the states of its neighbors as well as its own action, satisfying the following form in discrete time~\cite{mozafari2012oscillator}:

\begin{align*}
    %\label{eq:kuramoto_dynamics_discrete}
    \theta_i(t\!+\!1) \!=\! \theta_i(t) \!+\! dt \!\underbrace{\left(\!\omega_i(t) \!+\! a_i(t)\!+\!\left(\!\sum_{j \in N_i}\! K_{ij} \sin(\theta_j \!-\! \theta_i)\! \right)\!\!\right)}_{:= \dot\theta_i(t)}\! + \epsilon_i(t).
\end{align*}
Above, $\omega_i$ denotes the natural frequency of agent $i$, $dt$ is the discretization time-step, $K_{ij}$ denotes the coupling strength between agents $i$ and $j$, $a_i(t)$ is the action of agent $i$ at time $t$, and $\epsilon_i(t) \sim N(0,\sigma^2)$ is a noise term faced by agent $i$ at time $t$. We note that this fits into the localized transition considered in network MDPs. For the reward, we consider frequency synchronization to a fixed target $\omega_{\mathrm{target}}$. In this case, the local reward of each agent can be described as $r_i(\theta_{N_i},a_i) =  - \left|\dot\theta_i - \omega_{\mathrm{target}}\right|$.

\end{example}

To provide context for what follows, we review a few key concepts in RL. First, fixing a localized policy tuple \( \pi = (\pi_1, \ldots, \pi_n) \), the Q-function for this policy \( \pi \) is:
\begin{align*}
Q^{\pi}(s, a) \! = \!\! & \ \frac{1}{n} \sum_{i=1}^{n} \mathbb{E}_{a(t) \sim \pi(\cdot|s(t))} \left[ \sum_{t=0}^{\infty} \gamma^t r_i(s_i(t), a_i(t)) |\! s(0) = s, \!a(0) = a \right] \\
:= \!\! & \  \frac{1}{n} \sum_{i=1}^{n} Q_i^{\pi}(s, a).
\end{align*}
\normalsize
% \runyu{Add specification that $\pi$ is local policy}
% \vspace{-1mm}
In the last step, we defined the local $Q$-functions \( Q_i^{\pi}(s, a) \) which represent the $Q$ functions for the individual reward \( r_i \). Correspondingly, we can also define the local value function $V_i^\pi(s) = \int_{a} \pi(a \mid s) Q_i^\pi(s,a) da.$ We note that the global $Q(s,a)$ function can be obtained by averaging $n$ local $Q_i(s,a)$ functions. This plays an important role due to the policy gradient theorem, which states that the policy gradient can be computed with knowledge of the $Q(s,a)$ function.
% \bo{no need a lemma for PG. }
\begin{fact}[\cite{sutton1999policy}]
    \label{lemma:policy_gradient}
    Let \(d^\theta(s) = (1 - \gamma) \sum_{t=0}^{\infty} \gamma^t\textup{Pr}(s_t = s)\) %d_t^\theta(s)\), where \(d_t^\theta\) is the distribution of \(s(t)\) under a fixed policy \(\pi^{\theta}\) when \(s(0)\) is drawn from \(\mu_0\).
Then, we have

\[
\nabla_\theta J(\pi^{\theta}) = \mathbb{E}_{s \sim d^\theta, a \sim \pi^\theta(\cdot | s)}
\left[ Q^{\pi^\theta}(s, a) \nabla \log \pi^\theta(a | s) \right].
\]
\end{fact}

% \bo{this section is way too long, make the difficulties simple and straight. use itemize bullet point. } \zhaolin{I shortened this part by about 100 words.}
A natural approach to learning the $Q(s,a)$ function in the networked case is for each agent to learn its local $Q_i(s,a)$ function and share it across the network to form a global average. However, this poses a significant challenge when (i) the network size $n$ is large, and (ii) the individual state and action spaces $\gS_i$ and $\gA_i$ are continuous. Even if $\gS_i$ and $\gA_i$ are finite, representing $Q_i(s,a)$ requires $(\abs*{\gS_i} \times \abs*{\gA_i})^n$ entries, which grows exponentially with $n$. This challenge worsens with continuous spaces, which have infinite cardinality. To address this, we first explore the \emph{exponential decay property} from prior work, which improves scalability with network size. We then present our main contribution: integrating the exponential decay property with spectral representations from single-agent RL to derive scalable local $Q_i$-value function representations for continuous state-action network MDPs. We begin by discussing the exponential decay property.

% \subsection{Exponential decay property}

% \zhaolin{reduce discussion about exp decay property. after exp decay property, add one paragraph about problem statement. (restate challenge, for large SA, etc etc)}

\textbf{Exponential decay property.} ~
% Previous works have focused on achieving scalable networked control by approximating the local $Q_i^\pi(s,a)$ with a truncated $\hat{Q}_i^\pi(s_{N_i^\kappa},a_{N_i^\kappa})$ that depends only on states and actions within a $\kappa$-hop neighborhood of agent $i$, and hence is easier to represent, at least in the finite state-action space setting, since each agent only needs to keep track of $(\abs*{\gS_i} \times \abs*{\gA_i})^\kappa$ entries. 
The exponential decay property~\cite{pmlr-v120-qu20a,qu2020scalable,lin2021multi} is defined as follows.

\begin{definition}
\label{definition:c_rho_property}
    Given any $c > 0$ and $0 < \rho < 1$, the \( (c, \rho) \)-exponential decay property holds for a policy $\pi$ if given any natural number $\kappa$, for any \( i \in \mathcal{N}, s_{N_i^\kappa} \in \mathcal{S}_{N_i^\kappa}, s_{N_{-i}^\kappa} \in \mathcal{S}_{N_{-i}^\kappa}, a_{N_i^\kappa} \in \mathcal{A}_{N_i^\kappa}, a_{N_{-i}^\kappa} \in \mathcal{A}_{N_{-i}^\kappa} \), the local value function \( Q_i^{\pi} \) satisfies,

\[
\left| Q_i^{\pi}(s_{N_i^\kappa}, s_{N_{-i}^\kappa}, a_{N_i^\kappa}, a_{N_{-i}^\kappa}) - Q_i^{\pi}(s_{N_i^\kappa}, s'_{N_{-i}^\kappa}, a_{N_i^\kappa}, a'_{N_{-i}^\kappa}) \right| \leq c \rho^{\kappa+1}.
\]
As an immediate corollary, it follows that 

\[
\left| V_i^{\pi}(s_{N_i^\kappa}, s_{N_{-i}^\kappa}) - V_i^{\pi}(s_{N_i^\kappa}, s'_{N_{-i}^\kappa}) \right| \leq c \rho^{\kappa+1}.
\]
\end{definition}
We defer discussion about when the exponential decay property holds to Appendix~\ref{appendix:exponential_decay}. The power of the exponential decay property is that it immediately guarantees that the dependence of \( Q_i^{\pi} \) on other agents shrinks quickly as the distance between them grows, such that the true local $Q_i(s,a)$-functions can be approximated by truncated $\hat{Q}_i(s_{N_i^\kappa},a_{N_i^\kappa})$-functions up to an error that decays exponentially with $\kappa$. The truncated $\hat{Q}_i$ function is significantly easier to represent in the finite state-action setting since each agent only needs to keep track of $(\abs*{\gS_i} \times \abs*{\gA_i})^\kappa$ entries. However, continuous state and action space problems still pose a significant challenge. To overcome this, we will use the idea of spectral representations from linear MDPs and show how this can be adapted to the networked setting to yield truncated functions.

% \input{analysis}
%%%%%%%%%%%%%%%%%%%%%%%%%%%%%%%%%%%%%%%%%%%%%%%%%%%%%%%%%%

\section{Spectral representations for truncated approximations of local $Q_i$-value functions}\label{sec:spectral}
%%%%%%%%%%%%%%%%%%%%%%%%%%%%%%%%%%%%%%%%%%%%%%%%%%%%%%%%%%

% \bo{I suggest to start with spectral decomposition, which is more general. Then, link it to linear MDP with low-rank assumption as a remark. Meanwhile, directly work with network MDP, the connection to linear mdp can be specified as a remark. }

% \bo{keep in mind "linear MDP" is with finite low-rank. If here talk about exponential decay spectrum, please avoid the usage of linear MDP, which diminishes the novelty of the method. }
% \runyu{Maybe we can add some quick summary and restate the key question and contribution here. To emphasize that the spectral representation for truncated Q is our contribution, not just stating results in literature. Something like:
% As mentioned in the introduction, the central question we aim to address is how to identify an appropriate representation for large state-action network MDPs. In this section, we tackle this question by demonstrating that the spectral representation of local transition kernels provides an effective representation for the local $Q_i$-value functions.
% (Lemma \ref{lemma:Q_i_linear_decomp})}

To recap, the key question we face is this: how can we derive scalable local $Q_i$-value function representations in network problems with continuous state-action spaces, and integrate them into a scalable control framework? This forms the main contribution of our work. In this section, we tackle this question by demonstrating that the spectral representation of local transition kernels provides an effective representation for the local $Q_i$-value functions (see Lemma \ref{lemma:Q_i_linear_decomp} below).

We first motivate our analysis by reviewing representation learning in centralized RL via \emph{spectral decompositions}~\cite{jin2020provably,ren2022spectral}. From such works, we know that if the global $P(s' \mid s,a)$ admits a linear decomposition in terms of some spectral features $\phi(s,a)$ and $\mu(s')$, then the  $Q(s,a)$-value function can be linearly represented in terms of the spectral features $\phi(s,a)$. In the case of representing local $Q_i$-functions, this property can be stated as follows.

\begin{lemma}[Representing local $Q_i$-value functions via spectral decomposition of $P$ (cf.~\cite{jin2020provably})]
    Suppose the probability transition $P(s' \mid s,a)$ of the next state $s'$ given the current $(s,a)$ pair can be linearly decomposed as $P(s' \mid s,a) = \phi(s,a)^\top \mu(s')$ for some features $\phi(s,a) \in \bbR^D$ and $\mu(s') \in \bbR^D$, which we also refer to as spectral representations. Then, the local $Q_i$-value function admits the linear representation
    \begin{align*}
        Q_i^\pi(s,a) = \tilde{\phi}_i(s,a)^\top w_i^\pi,
    \end{align*}
    where
    \begin{align*}
        \tilde{\phi}_i(s,a) := [r_i(s_i,a_i), \phi(s,a)], \quad w_i^\pi = [1,\gamma \int_{s'} \mu(s') V_i^\pi(s') ds']^\top.
    \end{align*}
\end{lemma}
The benefit of the spectral decomposition property is that the $Q_i$-value functions can be represented by a $(D+1)$-dimensional representation $\tilde{\phi}_i(s,a)$ comprising the spectral representation $\phi(s,a) \in \bbR^{D}$ and local reward $r_i(s_i,a_i) \in \bbR$. However, applying this result directly in the networked case poses significant challenges, since the required feature dimension $D+1$ may be high.
% In the network setting, $s = (s_1,\dots,s_n) \in \bbR^{Sn}$ and $a = (a_1,\dots,a_n) \in \bbR^{An}$. 
To see why this is the case, recall that the probability transition in our networked setting admits the following factorization:
\begin{align*}
   \hspace{25pt}\textstyle \bbP(s' \mid s,a) = \prod_{i=1}^n \bbP(s_i' \mid s_{N_i},a_{N_i}).
\end{align*}
Assume each agent's transition probability has the following $d$-dimensional spectral decomposition.\begin{property}
    \label{property:spectral}
For any $i \in [n]$ and any state-action-next state tuple $(s,a,s')$, there exist features $\bar{\phi}_i(s_{N_i},a_{N_i}) \in \mathbb{R}^d$ and $\bar{\mu}_i(s_i') \in \mathbb{R}^d$ such that
\begin{align*}
\textstyle \prob{s_i'\mid s_{N_i}, a_{N_i}} = \langle \bar{\phi}_i(s_{N_i},a_{N_i}), \bar{\mu}_i(s_i')\rangle
\end{align*}
\end{property}
Given the factorization of the dynamics, this implies that 
\begin{align*}
    &\hspace{15pt} \ \textstyle\prob{s' \mid s, a} = \prod_{i=1}^n \langle \bar{\phi}_i(s_{N_i},a_{N_i}), \bar{\mu}_i(s_i')\rangle  \\  = &\textstyle \ \inner{\bigotimes_{i=1}^n \bar{\phi}_{i}(s_{N_i},a_{N_i})}{\bigotimes_{i =1}^n \bar{\mu}_i(s_i')}
    :=   \inner{\bar{\phi}(s,a)}{\bar{\mu}(s')}.
\end{align*}
Agnostically, this means that representing the global network dynamics may require using the $d^n$-dimensional features $\bar{\phi}(s,a) := \bigotimes_{i=1}^n \bar{\phi}_{i}(s_{N_i},a_{N_i})$, which even for small $d$ is undesirable due to an exponential dependence on the network size $n$. 

While the exponential decay property suggests that the $\hat{Q}_i$-function can be approximated by considering a $\kappa$-hop neighborhood of agent $i$, it is unclear how we can combine this with the spectral decomposition property to derive scalable representations for the local $Q_i$-value functions. 

To resolve this, we combine insights from both the exponential decay and spectral decomposition property, which intuitively, suggests that what matters in determining $Q_i^\pi(s,a)$ is the probability transition dynamics within a $\kappa$-hop neighborhood of agent $i$. In fact, due to the local factorization property of the dynamics, the evolution of $\kappa$-hop neighborhood only depends on the $\kappa+1$-hop neighborhood, which, when Property \ref{property:spectral} holds, admits the following spectral decomposition.

% \zhaolin{Intuitive, but how to do? make it sound non-trivial.}

\begin{property}[Network $\kappa$-local spectral features]
    \label{property:spectral_kappa}
For any $i \in [n]$ and any state-action-next state tuple $(s,a,s')$, there exist some positive integer $d_{i,\kappa}$ and features $\phi_{i,\kappa}(s_{N_i^{\kappa+1}},a_{N_i^{\kappa+1}}) \in \mathbb{R}^{d_{i,\kappa}}$ and $\mu_{i,\kappa}(s_{N_i^\kappa}') \in \mathbb{R}^{d_{i,\kappa}}$ such that
\begin{align*}
    \prob{s_{N_i^\kappa}' \mid s_{N_i^{\kappa+1}}, a_{N_i^{\kappa+1}}} = \langle \phi_{i,\kappa}(s_{N_i^{\kappa+1}},a_{N_i^{\kappa+1}}), \mu_{i,\kappa}(s_{N_i^\kappa}')\rangle.
\end{align*}
\end{property}

As shown in Lemma \ref{lemma:property1_implies_property2}, when Property \ref{property:spectral} is true, Property \ref{property:spectral_kappa} also holds, with $\phi_{i,\kappa}$ and $\mu_{i,\kappa}$ given by appropriate tensor products of the original $\bar{\phi}_i$ and $\bar{\mu}_i$ from the factorization of the local dynamics. We defer the proof to Appendix \ref{appendix:helper_factorization_prop1_implies_prop2}.
\begin{restatable}{lemma}{propOneImpliesPropTwo}
    \label{lemma:property1_implies_property2}
    Property \ref{property:spectral_kappa} holds whenever Property \ref{property:spectral} holds, by setting
    \begin{align*}
% \label{eq:tensor_product_rep}
    & \textstyle \ \phi_{i,\kappa}(s_{N_i^{\kappa+1}}(t), a_{N_i^{\kappa+1}}(t)) := \bigotimes_{j \in N_i^\kappa}  \bar{\phi}_j(s_{N_j}(t),a_{N_j}(t)), \\ & \textstyle \ \mu_{i,\kappa}(s_{N_i^{\kappa}}(t+1)) := \bigotimes_{j \in N_i^\kappa} \bar{\mu}_j(s_j(t+1)).
\end{align*}
\end{restatable}
\begin{remark}
    While the tensor product representation in Lemma \ref{lemma:property1_implies_property2} can be used to give a factorization of the $\kappa$-transition dynamics in Property \ref{property:spectral_kappa}, for specific problems, there may exist problem-specific alternative $\phi_{i,\kappa}$ and $\mu_{i,\kappa}$ features that may be lower-dimensional and thus more tractable to use.
\end{remark}

Property \ref{property:spectral_kappa} presents us with a path towards scalable representation of the local $Q_i$ via factorization of the local $\kappa$-hop neighborhood dynamics and approximating $Q_i(s,a)$ by network local representations. We first formalize this in the case when the spectral decomposition is exact and error-free. When this holds, we have the following lemma which shows how $Q_i^\pi(s,a)$ can be approximated by network local representations. We defer the proof to Appendix \ref{appendix:Qi_approx_error_exact}.
\begin{restatable}[Local $Q_i$ approximation via network $\kappa$-local spectral features]{lemma}{QiLinearDecomp}
    \label{lemma:Q_i_linear_decomp}
    Suppose the $(c,\rho)$-exponential decay property holds. Suppose Property \ref{property:spectral_kappa} also holds. Then, for any $(s,a)$ pair, agent $i$, and natural number $\kappa$, there exists an approximation $\bar{Q}_i^\pi$ which depends linearly on network $\kappa$-local spectral features $\tilde{\phi}_{i,\kappa}(s_{N_{i}^{\kappa+1}}, a_{N_{i}^{\kappa+1}})$, such that

    \begin{align*}
        \abs*{Q_i^\pi(s_{N_i^{\kappa+1}},a_{N_i^{\kappa+1}}, s_{N_{-i}^{\kappa+1}}, a_{N_i^{\kappa+1}}) - \bar{Q}_i^\pi(s_{N_i^{\kappa+1}},a_{N_i^{\kappa+1}})} \leq 2c \gamma \rho^{\kappa+1},
    \end{align*}
    \normalsize
    where $\bar{Q}_i^\pi(s_{N_i^{\kappa\!+\!1}},a_{N_i^{\kappa\!+\!1}})\! =\! \inner{\tilde{\phi}_{i,\kappa}(s_{N_i^{\kappa\!+\!1}},a_{N_i^{\kappa\!+\!1}})}{w_{i,\kappa}^\pi(s_{N_i^{\kappa}}^\prime)}$,\\
    \normalsize
with the definitions
\begin{align*}
    & \ \tilde{\phi}_{i,\kappa}(s_{N_i^{\kappa+1}},a_{N_i^{\kappa+1}}) := [r_i(s_i,a_i), \phi_{i,\kappa}(s_{N_i^{\kappa+1}}, a_{N_i^{\kappa+1}})]^\top, \\ 
    & \ w_{i,\kappa}^\pi(s_{N_i^{\kappa}}^\prime) := [1,\gamma \int_{s_{N_i^{\kappa}}^\prime}  ds_{N_i^\kappa}^\prime \mu_{i,\kappa}(s_{N_i^\kappa}^\prime) \bar{V}_i^\pi(s_{N_i^{\kappa}}^\prime) ]^\top,
\end{align*}
where 
$$\textstyle\hspace{20pt}\bar{V}_i^\pi(s_{N_i^{\kappa}}^\prime) := \int_{s_{N_{-i}^{\kappa}}^\prime} \frac{d s_{N_{-i}^{\kappa}}^\prime}{\mathrm{Vol}(\mathcal{S}_{N_{-i}^\kappa})} V_i^\pi(s_{N_i^{\kappa}}^\prime, s_{{N_{-i}^\kappa}}^\prime).$$
\end{restatable}

% \runyu{It would be better if we have a paragraph title, say 'Approximation' here?}
\textbf{Approximation.} In general, it may be impossible to find $\phi_{i,\kappa}$ and $\mu_{i,\kappa}$ that can exactly factorize the transition kernel. However, in both the unknown-model and the known-model cases, there exist ways to approximate the kernel. In the model-free case, we may leverage representation-learning techniques to approximate the spectrum of the $\kappa$-hop transition kernel, such as the spectral decomposition in \cite{ren2022spectral} which seeks to approximate the SVD of the kernel. In the model-based case, in the case when the local transition evolves according to a known dynamics function subject to Gaussian noise, we may approximate the kernel by random or Nystrom features ~\cite{ren_stochastic_2023}. We provide below a unified analysis for the error bound of approximating $Q_i(s,a)$ in terms of network $\kappa$-local representations $\hat{\phi}_{i,\kappa}(s_{N_i^{\kappa+1}}, a_{N_i^{\kappa+1}})$ that approximately factorize $P(s_{N_i^{\kappa}} \mid s_{N_i^{\kappa+1}}, a_{N_i^{\kappa+1}})$.
\begin{lemma}
\label{lemma:approx_error_from_approx_P}
    For any distribution $\nu^o$ over the space $\gS_{N_i^{\kappa+1}} \times \gA_{N_i^{\kappa+1}}$, suppose there exists a network $\kappa$-local representation $\hat{\phi}_{i,\kappa}(s_{N_i^{\kappa+1}}, a_{N_i^{\kappa+1}}) \in \bbR^m$ for which there exists $\hat{\mu}(s_{N_i^{\kappa}}^\prime) \in \bbR^m$ such that for every $i \in [n]$, the following holds for some approximation error $\ep_P > 0$:
\begin{align}
        \label{eq:phi_SVD_error_bound}
    \xpt[\nu^o]{\int_{s^\prime_{N_i^{\kappa}}}\!\! \left|P(s^\prime_{N_i^{\kappa}} \!\!\mid\!\! s_{N_i^{\kappa+1}}, a_{N_i^{\kappa+1}}) \!-\! \hat{\phi}_{i,\kappa}(s_{N_i^{\kappa+1}}, a_{N_i^{\kappa+1}})^\top \hat{\mu}_{i,\kappa}(s^\prime_{N_i^{\kappa}})\right| ds^\prime_{N_i^{\kappa}}} \leq \ep_P,
    % \resizebox{0.99\hsize}{!}{
    % $\xpt[\nu]{\int_{s^\prime_{N_i^{\kappa}}}}$}
\end{align}
Then, by setting $\tilde{\phi}_{i,\kappa}(s_{N_i^{\kappa+1}}, a_{N_i^{\kappa+1}}) := [r_i(s_i,a_i), \hat{\phi}_{i,\kappa}(s_{N_i^{\kappa+1}}, a_{N_i^{\kappa+1}})]^\top$, for every $i \in [n]$,
    \begin{align}
    \label{eq:phi_approx_err_bounded_by_svd_error}
        \min\limits_{w \in \bbR^{m+1}}  \xpt[\nu^o] {\abs*{\bar{Q}_i^\pi(s_{N_i^{\kappa+1}}, a_{N_i^{\kappa+1}}) - \tilde{\phi}_{i,\kappa}(s_{N_i^{\kappa+1}}, a_{N_i^{\kappa+1}})^\top w}} \leq \frac{\ep_P \gamma \bar{r}}{1-\gamma}.
    \end{align}
\end{lemma}
    \begin{proof}
        Suppose (\ref{eq:phi_SVD_error_bound}) holds. Then, define $w^* := [1,\gamma \int_{\gS_{N_i^\kappa}} \hat{\mu}_{i,\kappa}(s^\prime_{N_i^\kappa}) \bar{V}_{i}^\pi(s_{N_i^\kappa}^\prime) ds_{N_i^\kappa}^\prime]^\top \in \bbR^{m+1}$. Then, by using the upper bound $\abs*{\bar{V}_{i}^\pi(s_{N_i^\kappa}^\prime)} \leq \frac{\bar{r}}{1-\gamma}$, we have 
        \begin{equation*}
        \min\limits_{w \in \bbR^{m+1}} \xpt[\nu]{\abs*{\bar{Q}_i^\pi(s_{N_i^{\kappa+1}}, a_{N_i^{\kappa+1}}) - \tilde{\phi}_{i,\kappa}(s_{N_i^{\kappa+1}}, a_{N_i^{\kappa+1}})^\top w}}\leq \frac{\ep_P \gamma \bar{r}}{1-\gamma}.
    \end{equation*}
    \end{proof}

The approximation error in the bound above relies on the condition in (\ref{eq:phi_SVD_error_bound}) to hold. In the case when the local transition evolves according to a known dynamics function subject to a positive-definite kernel noise (e.g. Gaussian noise), we may approximate the $\kappa$-hop transition kernel with random features such that (\ref{eq:phi_SVD_error_bound}) holds with high probability. For clarity of exposition, we focus on the approximation error of random features for Gaussian kernels~\cite{rahimi2007random}\footnote{We note that our result easily generalizes to any positive-definite transition kernel noise (e.g. Laplacian, Cauchy, Mat\'ern, etc; see Table 1 in \cite{dai2014scalable} for more examples)}. In this case, our error bound is shown in the following result, whose proof we defer to Appendix \ref{appendix:random_features_approx_error}.
\begin{restatable}{lemma}{rfApproxError}
\label{lemma:spectral_decomp_gaussian}
Fix any $i \in [n]$. Suppose the local dynamics take the form 
$s_i^\prime = f_i(s_{N_i},a_{N_i}) + \epsilon_i$ where $\epsilon_i \sim N(0,\sigma^2 I_{S})$, such that for any $\kappa$, $s_{N_i^\kappa}^\prime = f_{i,\kappa}(s_{N_i^{\kappa+1}},a_{N_i^{\kappa+1}}) + \epsilon_{N_i^\kappa}$ where $\epsilon_{N_i^\kappa} \sim N(0,\sigma^2 I_{\abs*{N_i^\kappa} S})$ and $f_{i,\kappa}$ is concatenation of $f_j$ for $j \in N_i^\kappa$. Fix any $0 \leq \alpha < 1$. Then, for a positive integer $m$, define the $m$-dimensional features $\hat{\phi}_{i, \kappa}(s_{N_i^{\kappa+1}},a_{N_i^{\kappa+1}}) \in \bbR^{m}$, where 

\begin{align*}
& \ \hat{\phi}_{i,\kappa}(s_{N_i^{\kappa+1}},a_{N_i^{\kappa+1}}) :=  \frac{g_\alpha(s_{N_i^{\kappa+1}},a_{N_i^{\kappa+1}})}{\alpha^{\abs*{N_i^\kappa}S}} \left\{\sqrt{\frac{2}{m}}\cos\left(\frac{\omega_\ell^\top f_{i,\kappa}(s_{N_i^{\kappa+1}},a_{N_i^{\kappa+1}})}{\sqrt{1-\alpha^2}} + b_\ell\right) \right\}_{\ell=1}^m,
\end{align*}
\normalsize
with $\{\omega_\ell\}_{\ell=1}^m$ being i.i.d draws from $N(0, \sigma^{-2}I_{\abs*{N_i^\kappa} S})$, $\{b_\ell\}_{\ell=1}^m$ being i.i.d draws from $\mathrm{Unif}([0,2\pi])$, and $g_\alpha(s_{N_i^{\kappa+1}},a_{N_i^{\kappa+1}}) := \exp\left(\frac{\alpha^2\norm*{f_{i,\kappa}(s_{N_i^{\kappa+1}},a_{N_i^{\kappa+1}})}^2}{2(1-\alpha^2)\sigma^2} \right)$. Define $\tilde{g}_\alpha := \max_{i \in [n]}\left(\sup_{x \in f_{i,\kappa}(\gS_{N_i^{\kappa+1}},\gA_{N_i^{\kappa+1}})} \frac{g_\alpha(x)}{\alpha^{\abs*{N_i^\kappa}S}}\right)$. Suppose $m = \Omega\left(\max\limits_{i \in [n]}\left[\log\left(\frac{\left(\abs*{N_i^\kappa}S (diam(\gS_{N_i^\kappa}))^2 \right)}{\sigma^2(\delta/n) (\epsilon_P/\tilde{g}_\alpha)} \right)\frac{\abs*{N_i^\kappa}S \tilde{g}_\alpha^2}{\ep_P^2}\right]\right)$ for some $\delta > 0$. Then, with probability at least $1 - \delta$, the condition in (\ref{eq:phi_SVD_error_bound}) holds for every $i \in [n]$ and any distribution $\nu^o$ over $\gS \times \gA$, with 
\begin{align*}
\hat{\mu}_{i,\kappa}(s^\prime_{N_i^{\kappa}})  \!:=\! \left\{\sqrt{\frac{2 }{m}}p_\alpha(s^\prime_{N_i^\kappa})\cos(\sqrt{1-\alpha^2}\omega_\ell^\top s^\prime_{N_i^{\kappa}}\! +\! b_\ell)\right\}_{\ell=1}^m\!,
\end{align*}
\normalsize
where $p_\alpha(s^\prime_{N_i^\kappa}):= \frac{\alpha^{\abs*{N_i^\kappa}S}}{(2\pi\sigma^2)^{\abs*{N_i^\kappa}S}} \exp(-\frac{\norm*{\alpha s^\prime_{N_i^\kappa}}^2}{2\sigma^2})$ is a Gaussian distribution with standard deviation $\frac{\sigma}{\alpha}$.
\end{restatable}
The key takeaway from the above result is that under Gaussian noise and known reward and dynamics function, there exists finite-dimensional features that can, with high probability, approximately factorize the local $\kappa$-transition kernels, satisfying the condition in (\ref{eq:phi_SVD_error_bound}) with high probability. Moreover, from this result, we note that the required number of features to achieve this is $\tilde{O}\left( \frac{\max_{i \in [n]}\abs*{N_i^\kappa}S \tilde{g}_\alpha^2}{\ep_P^2}\right)$, which only depends on the dimension of states in the largest $\kappa$-hop neighborhood. We note that the tunable $\alpha$ in Lemma \ref{lemma:spectral_decomp_gaussian} allows greater flexibility and may be tuned to improve empirical performance~\cite{ren_stochastic_2023}.

%%%%%%%%%%%%%%%%%%%%%%%%%%%%%%%%%%%%%%%%%%%%%%%%%%%%%%%%%%

\section{Algorithms}\label{sec:algorithm}
%%%%%%%%%%%%%%%%%%%%%%%%%%%%%%%%%%%%%%%%%%%%%%%%%%%%%%%%%%

As suggested in Lemma \ref{lemma:Q_i_linear_decomp}, $\tilde\phi_{i,\kappa}$ serves as a good representation for the local $Q_i$-functions. Based upon this observation, this section ocuses on how the local $Q_i$-function and subsequently a good localized policy can be learned. The algorithm contains three major steps: \textbf{feature generation},  \textbf{policy evaluation} and  \textbf{policy gradient}.

The first step is \textbf{feature generation} (Lines \ref{algline:spectral_embedding} through \ref{algline:end_spectral_embedding}), where we generate the appropriate features $\tilde\phi_{i,\kappa}$. This comprises the local reward function as well as the spectral features $\hat{\phi}_{i,\kappa}(s_{N_i^{\kappa}, a_{N_i^{\kappa}}})$ coming from the factorization of the local $\kappa$-hop dynamics. In the case of known dynamics and Gaussian noise, we know from Lemma \ref{lemma:spectral_decomp_gaussian} that $\hat{\phi}_{i,\kappa}(s_{N_i^{\kappa}, a_{N_i^{\kappa}}})$ can be derived by random features which factorize the local $\kappa$-hop dynamics with high probability. In this case, we note that our spectral features are scalable with respect to both the network size and the continuous state-action space, since the required number of features only depend on the dimensions of the $\kappa$-hop neighborhoods.

The second step is \textbf{policy evaluation}, where we use the feature $\tilde\phi_{i,\kappa}$ and apply LSTD to find a set of weights $w_i$ to approximate the local $Q_i$-functions by $\hat Q_i = \widetilde\phi_{i,\kappa}^\top \hat w_i$. At each round $k \in [K]$, we first sample $M_s$ samples from the stationary distribution of $\pi^{(k)}$ (Line \ref{algline:data_sampling}), and then perform a LSTD update for each agent $i \in [n]$ to learn the appropriate weights for the local $Q_i$-functions (Line \ref{algline:policy_eval_start}). 
% \runyu{Shall we delete the following sentence to save space? It is a bit repetitive.} As we will see in Lemma \ref{lemma:policy_eval_error_main}, apart from a statistical error that decays as the number of samples $M_s$ increases, the policy evaluation error contains a truncation error that decays as $\kappa$ increases, and an approximation error that depends on the accuracy to which the features factorize the local $\kappa$-hop dynamics ($\ep_P$).

Finally, the last step is updating policy using \textbf{policy gradient} (Lines \ref{algline:policy_eval_start} to \ref{algline:policy_eval_end}). For each agent $i \in [n]$, with the learned $\{\hat{Q}_j\}_{j \in N_i^{\kappa_\pi + \kappa}}$, we perform a gradient step to update the local policy weights $\theta_i$, and update to the new policy. We note that this update is scalable since from the perspective of each agent $i$, it only requires knowledge of the local $\hat{Q}_j$ for agents $j$ in a $(\kappa_\pi + \kappa)$-hop neighborhood of agent $i$. In practice, the $\kappa$-hop spectral representation we introduce can be combined flexibly with any actor in a distributed cooperative actor-critic framework that requires knowledge of the local $Q_i$-functions.

\RestyleAlgo{ruled}
\SetKwComment{Comment}{}{}
\setlength{\algomargin}{1em} % Adjusts the left margin of the algorithm
\SetInd{0.3em}{0.2em} %
\begin{algorithm}
\LinesNumbered
\caption{Networked control with spectral embedding}\label{alg:networked_spectral_control}

\KwData{$Q$-value truncation radius $\kappa$,  Policy truncation radius $\kappa_\pi$, Reward Function $r(s, a)$, Number of features $m$, Number of samples/round $M_s$, Learning Rate $\eta$, Number of rounds K}
\KwResult{$\pi^{(K+1)} = (\pi_1^{(K+1)},\dots,\pi_n^{(K+1)})$}

\Comment{\color{blue}Spectral dynamic embedding generation}
\label{algline:spectral_embedding}
\For{$i \in [n]$}
{
Generate features $\tilde{\phi}_{i,\kappa}(s_{N_i^{\kappa+1}},a_{N_i^{\kappa+1}}) \!:= \![r_i(s_i,a_i), \hat{\phi}_{i,\kappa}(s_{N_i^{\kappa+1}},a_{N_i^{\kappa+1}})] \!\in\! \mathbb{R}^{m+1}$\!,\normalsize where $\hat{\phi}_{i,\kappa}(s_{N_i^{\kappa\!+\!1}}\!,a_{N_i^{\kappa\!+\!1}})$ \normalsize  satisfies the condition in (\ref{eq:phi_SVD_error_bound}).
}
\label{algline:end_spectral_embedding}

\Comment{\color{blue}Policy evaluation and update}

% truncated spectral feature or random Fourier feature. 
\For{$k= 1,2, \cdots, K$}
{
    \Comment{{\color{blue}Least squares policy evaluation}}

    Set $\pi_i^{(k)} := \pi_i^{\theta_i^{(k+1)}}$. Sample \iid~ $D_k  = \{(s(j), a(j), s^\prime(j)), a^\prime(j)\}_{j \in [M_s]}$ where $(s(j), a(j)) \sim \nu_{\pi^{(k)}}$, $s^\prime(j) \sim P(\cdot \mid s(j), a(j))$ where $\nu_{\pi^{(k)}}$ is the stationary distribution of $\pi^{(k)}$, and
    $\forall j \in [M_s], \forall i \in [n]: a_i^\prime(j) \sim \pi_i^{(k)}(\cdot \mid s_{N_i^{\kappa_\pi}}^\prime(j))$
        \label{algline:data_sampling}
    \For{$ i \in [n]$}{
    %Initialize $\hat{w}_{i, 0}^{(k)} = 0$.
    Solve $\hat{w}_{i}^{(k)}$ using least square temporal difference (LSTD) as follows:
        \label{algline:policy_eval_start}
        \begin{align}
        \hat{w}_i^{(k)} &= (M_i^{(k)})^{-1} H_i^{(k)} r_i \notag\\
       M_i^{(k)}&\!:=\! \textstyle \frac{1}{|D_k|}\sum_{s,a,s'\!,a'\!\in D_k} \!\!\tilde{\varphi}_{i,\kappa} (\tilde{\varphi}_{i,\kappa} \!-\! \gamma\tilde{\varphi}'_{i,\kappa})^\top \label{eq:def-M_ik}\\
       H_i^{(k)}&\!:=\textstyle \!\frac{1}{|D_k|}\sum_{s,a,s',a'\in D_k} \tilde{\varphi}_{i,\kappa} \tilde{\varphi}_{i,\kappa}^\top\notag\\
       r_i &:= [1,0,0,\dots,0]^\top\in\bbR^{m +1}\notag
    \end{align}
    where 

    \begin{align*}
        &\textstyle \tilde\varphi_{i,\kappa}(j) := \tilde\phi_{i,\kappa}(s_{N_i^{\kappa+1}}(j), a_{N_i^{\kappa+1}}(j)), \\
        & \textstyle\tilde\varphi'_{i,\kappa}(j) := \tilde\phi_{i,\kappa}(s'_{N_i^{\kappa+1}}(j), a'_{N_i^{\kappa+1}}(j)).
    \end{align*}
    \normalsize
    % \For{$t=0, 1, \cdots, T-1$}
    % {
    %     Solve \runyu{Actually this is not LSTD, will re-write this part later}
    %     \begin{align}
    %     \resizebox{0.99\hsize}{!}{
    %     $\hat{w}_{i, t+1}^{(k)} = \mathop{\arg\min}\limits_{w} 
    %     \left\{ \sum\limits_{j=1}^{M_s} \left(\tilde\varphi_{i,\kappa}(j)^\top w - r(s_i, a_i) - \gamma \tilde\varphi'_{i,\kappa}(j))^\top \hat{w}_{i, t}^{(k)}\right)^2\right\},$
    %     }
    %     \label{eq:projected_least_square}
    %     \end{align}
    %     where \resizebox{0.8\hsize}{!}{$\tilde\varphi_{i,\kappa}(j) := \tilde\phi_{i,\kappa}(s_{N_i^{\kappa+1}}(j), a_{N_i^{\kappa+1}}(j)), \ \ \tilde\varphi'_{i,\kappa}(j) := \tilde\phi_{i,\kappa}(s'_{N_i^{\kappa+1}}(j), a'_{N_i^{\kappa+1}}(j)).$}
    % }
    Update approximate $\hat{Q}_i^{(k)}$-value function as $\hat{Q}_i(s_{N_i^{\kappa+1}}, a_{N_i^{\kappa+1}}) := \tilde\phi_{i,\kappa}(s_{N_i^{\kappa+1}}, a_{N_i^{\kappa+1}})^\top \hat{w}_{i}^{(k)}.$
    }
    \label{algline:policy_eval_end}
    \Comment{{\color{blue} Policy gradient for control}}
    \label{algline:policy_gradient_start}
    \For{$i \in [n]$} {
    Calculate 
    \begin{small}
    \begin{align*}
        \textstyle \hat{g}_i^{(k)} = \frac{1}{M_s}\sum\limits_{j =1}^{M_s}  \sum\limits_{\ell \in N_i^{\kappa +\kappa_\pi}} \frac{\hat{Q}_\ell^{(k)}(s_{N_\ell^{\kappa}}(j), a_{N_\ell^{\kappa}}(j)) }{n} \times \\
        \textstyle \nabla_{\theta_i}\log \pi_i^{(\theta_i^{(k)})}(a_i(j) \mid s_{N_i^{\kappa_\pi}}(j))
    \end{align*}
    \end{small}
    %\resizebox{0.999\hsize}{!}{$$.}
    \label{algline:gradient_calc}
    
    Take gradient step $\theta_i^{(k+1)} = \theta_i^{(m)} + \eta \hat{g}_i^{(k)}$
    
    }
}
\end{algorithm}

\subsection{Policy evaluation error}
We have the following result on the policy evaluation error with our network $\kappa$-local features. We defer the details of the proof (including preliminary results required for the proof) to Appendix \ref{appendix:policy_eval_error}.
\begin{lemma}[Policy Evaluation Error]
\label{lemma:policy_eval_error_main}
Suppose condition (\ref{eq:phi_SVD_error_bound}) in Lemma \ref{lemma:approx_error_from_approx_P} holds. Suppose the sample size $M_s \ge \log\left(\frac{2(m+1)}{\delta/(Kn)}\right)$. Then, with probability at least $1-2\delta$, for every $i \in [n]$ and $k \in [K]$, the ground truth $Q$ function $Q^{\pi(k)}_i(s,a)$ and the truncated $Q$ function learnt in Algorithm \ref{alg:networked_spectral_control} $\hat{Q}_i(s_{N_i^{\kappa+1}}, a_{N_i^{\kappa+1}})$ satisfies, for any distribution $\bar\nu$ on $\gS \times \gA$,
% \tiny
% \begin{align*}
%    & \ \left(\xpt[\bar\nu]{\left| Q^{\pi(k)}_i(s,a) - \hat{Q}_i^{(k)}(s_{N_i^{\kappa+1}}, a_{N_i^{\kappa+1}})\right|^\lambda}\right)^{\frac{1}{\lambda}} \\
%    \le & \ O\left(c\rho L^2 D\rho^{\kappa+1} +\log\left(\frac{(m+1)}{\delta/(Kn)}\right) \frac{D^2L^5}{\sqrt{M_s}} + L\ep_P \left( \norm*{\frac{\bar\nu}{\nu^o}}_\infty + \norm*{\frac{\nu_{\pi^{(k)}}}{\nu^o}}_\infty\right)\right),
% \end{align*}
% \normalsize

\begin{align*}
   & \ \xpt[\bar\nu]{\left| Q^{\pi(k)}_i(s,a) - \hat{Q}_i^{(k)}(s_{N_i^{\kappa+1}}, a_{N_i^{\kappa+1}})\right|} \\
   \le & \!\ O\!\left(\underbrace{c\rho L^2 D\rho^{\kappa+1}}_{\textup{truncation error}}\!\!+\underbrace{\log\left(\frac{(m\!+\!1)}{\delta/(Kn)}\right) \frac{D^2L^5}{\sqrt{M_s}}}_{\textup{statistical error}} + \underbrace{L\frac{\ep_P \gamma \bar{r}}{1-\gamma} \left( \norm*{\frac{\bar\nu}{\nu^o}}_\infty \!\!+ \norm*{\frac{\nu_{\pi^{(k)}}}{\nu^o}}_\infty\right)}_{\textup{approximation error}}\right),
\end{align*}
\normalsize
where %(denoting \resizebox{0.6\hsize}{!}{$\tilde{\varphi}_{i,\kappa} :=  \tilde{\phi}_{i,\kappa}(s_{N_i^{\kappa+1}}, a_{N_i^{\kappa+1}}), \tilde{\varphi}_{i,\kappa}' :=  \tilde{\phi}_{i,\kappa}(s'_{N_i^{\kappa+1}}, a'_{N_i^{\kappa+1}})$})

\begin{align*}
    & \hspace{10pt} D := \max_{i \in [n], k \in [K]} \norm*{(M_i^{(k)})^{-1}}, \quad L := \max_{i \in [n]} \norm*{\tilde{\varphi}_{i,\kappa}}
\end{align*}
and $M_i^{(k)}$ is defined as in \eqref{eq:def-M_ik}
\normalsize
\end{lemma}
From the above result, we note that the policy evaluation error comprises three components, with one being the statistical error due to using finite samples, which decays with the square root of the sample size $M_s$, and the truncation error from considering a truncated $\kappa$-hop neighborhood (this decays exponentially in $\kappa$), as well as the approximation error of the spectral features in approximating the $\kappa$-hop transition ($\ep_P$). 

\subsection{Policy optimization error and main convergence result}

\begin{theorem}
\label{theorem:main}
Suppose the sample size $M_s \ge \log\left(\frac{2(m+1)}{(\delta/Kn)}\right)$. Suppose with probability at least $1 - \delta$, for all $ i \in [n]$, the following holds for some features $\hat\phi_{i,\kappa} \in \bbR^m$ and $\hat\mu_{i,\kappa} \in \bbR^m$:
\begin{align*}
\xpt[\nu^o]{\int_{s^+_{N_i^{\kappa}}}\!\! \left| P(s^+_{N_i^{\kappa}} \!\!\mid\!\! s_{N_i^{\kappa+1}}, a_{N_i^{\kappa+1}}) \!-\! \hat{\phi}_{i,\kappa}(s_{N_i^{\kappa+1}}, a_{N_i^{\kappa+1}})^\top \hat{\mu}_{i,\kappa}(s^+_{N_i^{\kappa}})\right| ds^+_{N_i^{\kappa}}} \leq \ep_P
\end{align*}
for some $\ep_P > 0$ and distribution $\nu^o$ over $\gS \times \gA$. Then, if $\eta = O(1/\sqrt{K})$, with probability at least $1-4\delta$,  
\begin{align*}
\frac{1}{K}\sum_{k=1}^{K} \norm*{\nabla J(\theta^{(k)})}^2 \leq O\left(\frac{\bar{r}/(1-\gamma)}{ \sqrt{K}} + \frac{L_\pi\bar{r}\ep_J}{1-\gamma} + \frac{L'}{\sqrt{K}}\left(\ep_J^2 + \left(\frac{L_\pi \bar{r}}{1-\gamma}\right)^2\right) \right),
\end{align*}
where $\ep_J := 2cL_\pi \rho^\kappa + \frac{2\bar{r}L_{\pi}}{1-\gamma}\sqrt{\frac{1}{M_s} \log\left(\frac{d_\theta +1}{\delta/K} \right) } + \epsilon_Q L_\pi,$
and

\begin{align*}
\ep_Q :=  \max\limits_{k \in [K]}O\left(c\rho L^2 D\rho^{\kappa+1} +\log\left(\frac{(m+1)}{\delta/(Kn)}\right) \frac{D^2L^5}{\sqrt{M_s}} + L\frac{\ep_P \gamma \bar{r}}{1-\gamma} \left( \norm*{\frac{\hat\nu^{(k)}}{\nu^o}}_\infty + \norm*{\frac{\nu_{\pi^{(k)}}}{\nu^o}}_\infty\right)\right),
\end{align*}
where $L'$ is the Lipschitz continuity parameter of $\nabla J(\theta)$, $L_{i,\pi}$ is a bound on $\norm*{\nabla_{\theta_i}\log \pi_i^{\theta_i}(\cdot \mid \cdot)}$, and $L_\pi := \sqrt{\sum_{i=1}^n L_{i,\pi}^2}$.
\normalsize
\end{theorem}
From the above result,  we see that our algorithm can achieve convergence to an approximate stationary point of the global objective $J$ as the number of rounds $K$ increases, up to an error term depending on $\ep_J$, which depends on the policy evaluation error $\ep_Q$ from Lemma \ref{lemma:policy_eval_error_main}. As we observed before, the policy evaluation error comprises a statistical error, a truncation error decaying exponentially as $\kappa$ increases, and a feature approximation error term $\ep_P$. Consequently, the convergence error to an approximation stationary point also depends on these three terms. 
% We note that by making further assumptions on the spatially decaying property of our policy class, it may be possible to achieve convergence to an approximate global optimum of the objective $J$, akin to the result shown in ~\cite{zhang2023global} for finite state-action spaces. We leave this possible investigation to future work.
%%%%%%%%%%%%%%%%%%%%%%%%%%%%%%%%%%%%%%%%%%%%%%%%%%%%%%%%%%
 % \vspace{-0.5em}
\section{Simulations}\label{sec:simulations}
 % \vspace{-0.5em}
%%%%%%%%%%%%%%%%%%%%%%%%%%%%%%%%%%%%%%%%%%%%%%%%%%%%%%%%%%

\subsection{Thermal control of multi-zone buildings}
% \vspace{-5pt}
We consider a stochastic linear dynamical system modeling the thermal control of a 50-zone building. We assume that the network is connected, with each agent having 2 neighbors. The dynamics of each agent is only affected by its neighbors, and subject to Gaussian noise. We also assume access to the model dynamics and reward function. In this problem, to implement our algorithm, we utilize random features that factorize the $\kappa$-hop Gaussian transition (cf. Lemma \ref{lemma:spectral_decomp_gaussian}), and perform least squares, followed by normalized gradient descent. The controller is parameterized to be linear. More details on our experimental setup can be found in Appendix \ref{appendix:simulations}. 

Since the dynamics are assumed to be linear, we have access to the cost of the optimal controller, making this a good way to benchmark the performance of our algorithm. The performance of our algorithm is shown in Figure \ref{fig:hvac_linear} below. As we can see, as $\kappa_\pi$ increases, our algorithm is indeed able to approximate the performance of the optimal controller. Moreover, the speed at which it converges is faster than $\hat{Q}_i$ approximations that leverage a generic two-hidden layer neural network (NN) to represent the (truncated) local $\hat{Q}_i$ value functions; we note that in both cases, the algorithms utilize the same learning rate for the policy gradient step, and have access to the rewards and dynamics function.

\begin{figure}[h]  % Single-column figure environment
  \centering
  \begin{subfigure}{0.49\linewidth}  % Adjust the subfigure width to fit within a column
    \includegraphics[width=\linewidth]{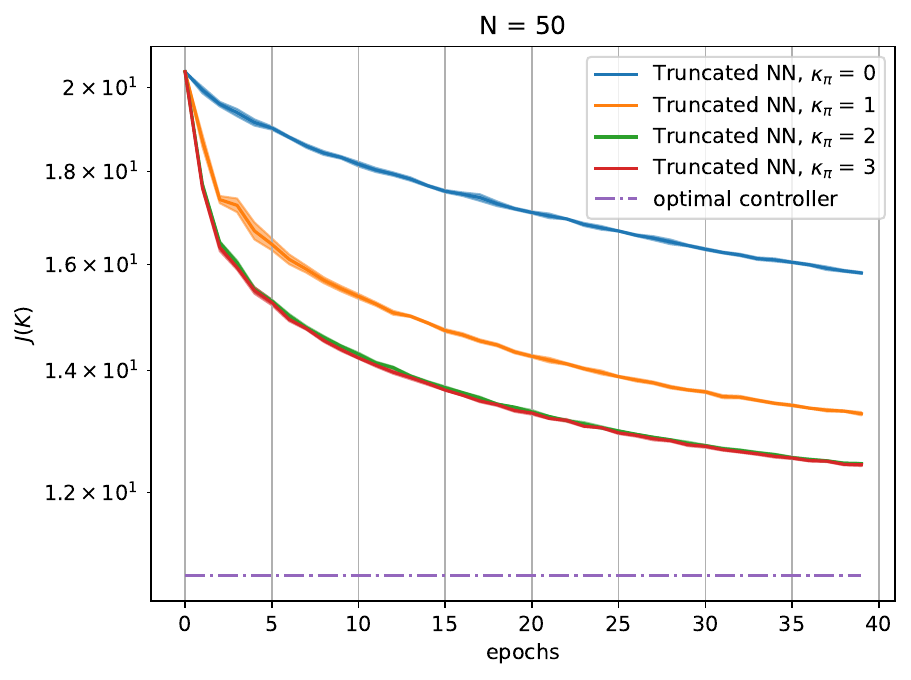}
    \caption{NN critic}
    \label{fig:hvac_nn}
  \end{subfigure}
  \hfill
  \begin{subfigure}{0.49\linewidth}  % Adjust the subfigure width to fit within a column
    \includegraphics[width=\linewidth]{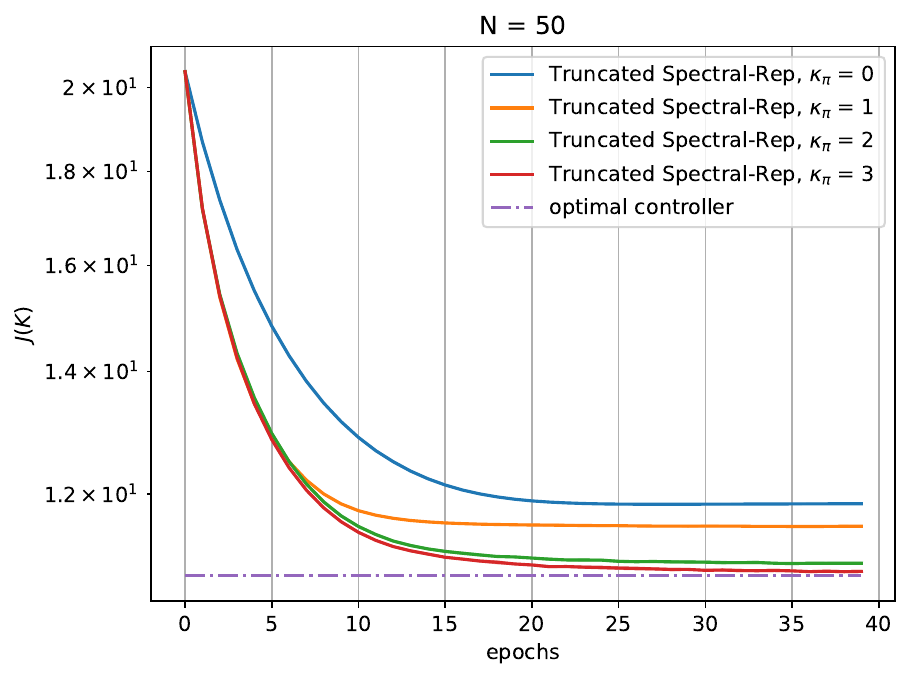}
    \caption{Spectral features critic}
    \label{fig:hvac_spectral}
  \end{subfigure}
  \caption{Learning trajectories of cost (lower is better) using Algorithm 1 + random features and NN critics on a 50-dimensional stochastic linear dynamical system for varying $\kappa_\pi$. Average and 1 std confidence intervals over 5 seeds.}
  \label{fig:hvac_linear}
  \vspace{-10pt}
\end{figure}

% \begin{figure}[h]
% \vspace{-10pt}
%   \centering
%   \includegraphics[width=0.7\hsize]{plots/learning_trajectory_N=50_trunc_kappa=3_kappa_comparison_m=800_n_data_eps=500_nepochs=40_batch_diff_keys.pdf}
%   \caption{Learning trajectory of cost (lower is better) using random features on a 50-dimensional stochastic linear dynamical system for varying $\kappa_\pi$.}
%   \label{fig:hvac_linear}
%   \vspace{-10pt}
% \end{figure}

% \zhaolin{Will add in the policy eval error comparison (between NN and random features later.}

\subsection{Kuramoto oscillator control}
\vspace{-5pt}
Due to the nonlinearity in this problem, we adopt the Soft-Actor-Critic (SAC) framework for this problem, and compare the performance of a generic deep NN critic with our network spectral local-$\kappa$ critic. In this problem, we consider the more realistic and difficult setting where the dynamics is unknown. In this problem, the network has 20 agents in total, and the network graph is connected, with each agent having 2 neighbors. We set the goal for the agents to synchronize to a target frequency of $0.75$.

In both the generic SAC and our spectral SAC implementation, the local critic for $Q_i$-value function considers a $\kappa$-hop neighborhood, i.e. approximate $Q_i$ by $\hat{Q}_i(s_{N_i^\kappa},a_{N_i^\kappa}) = \hat{\phi}_i(s_{N_i^\kappa},a_{N_i^\kappa})^\top w_i$, where $\hat{\phi}_i(s_{N_i^\kappa},a_{N_i^\kappa})$ is a two-hidden layer neural network. However, for our approach (spectral + SAC), we add a feature step that regularizes the feature $\hat{\phi}_i(s_{N_i^\kappa},a_{N_i^\kappa})$ towards factorizing the local dynamics, i.e. minimizing the objective in Condition \ref{eq:phi_SVD_error_bound} in Lemma \ref{lemma:approx_error_from_approx_P}. We defer more details on the problem setup as well as experimental details to Appendix~\ref{appendix:simulations}.

\begin{figure}[h]
  \centering
  \includegraphics[width=0.7\hsize]{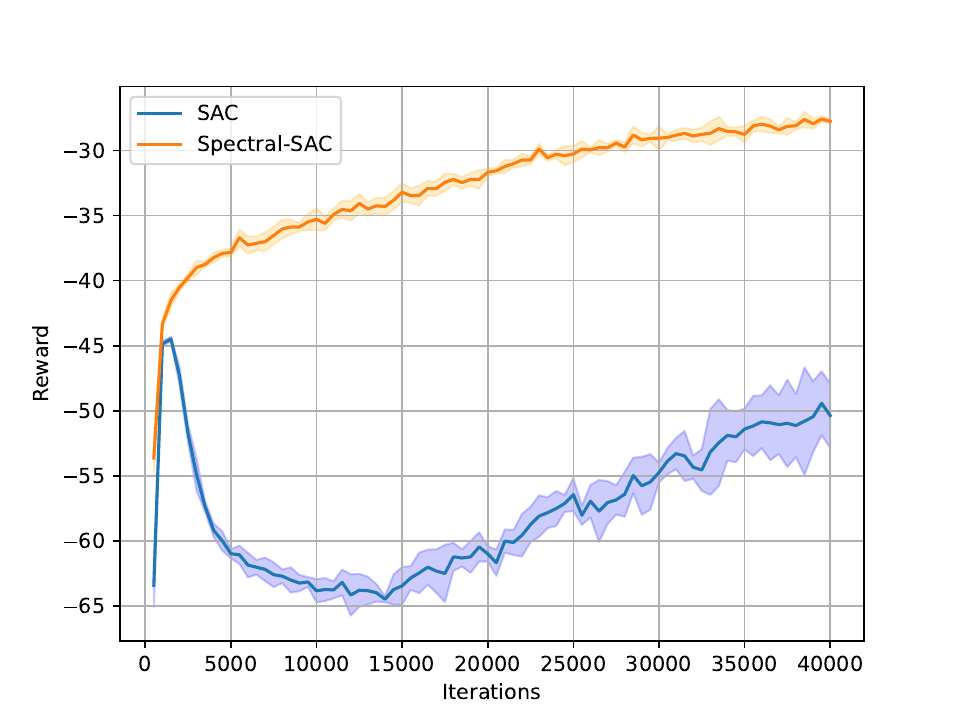}
  \caption{Change in reward during training for Kuramoto oscillator control, $n = 40$, $\kappa_\pi = 1, \kappa = 2$. The performance for each algorithm is averaged over 5 seeds.}
  \label{fig:kuramoto_learning}
  \vspace{-10pt}
\end{figure}

In Figure \ref{fig:kuramoto_learning}, we compare the performance of our approach (Spectral + SAC) with a generic SAC with two-hidden layer NN critic. As we can see, our approach leads to significantly higher rewards. Moreover, we observe that our approach tends to lead to qualitatively better synchronization behavior when starting from the same initial condition, as suggested in Figure \ref{fig:kuramoto_trajectory}. Finally, in Appendix~\ref{appendix:simulations}, we note that in the model-based setting, our algorithm (utilizing random features) achieves a performance comparable to that of generic NN approaches.

\vspace{-5pt}

\begin{figure}[h]  % Single-column figure environment
  \centering
  \begin{subfigure}{0.49\linewidth}  % Adjust the subfigure width to fit within a column
    \includegraphics[width=\linewidth]{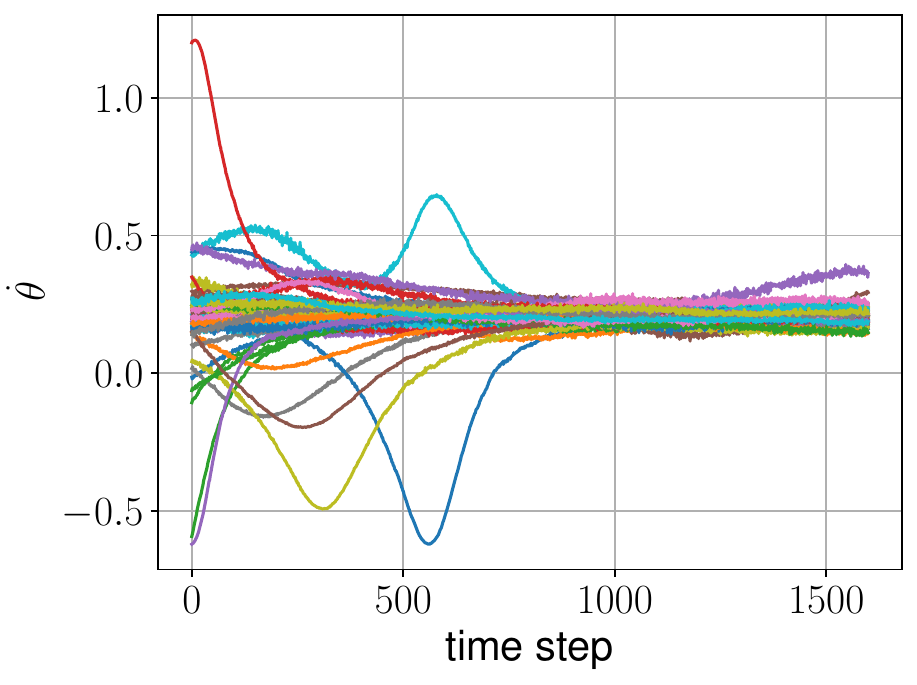}
    \caption{SAC Controller}
    \label{fig:kuramoto_controller_sac}
  \end{subfigure}
  \hfill
  \begin{subfigure}{0.49\linewidth}  % Adjust the subfigure width to fit within a column
    \includegraphics[width=\linewidth]{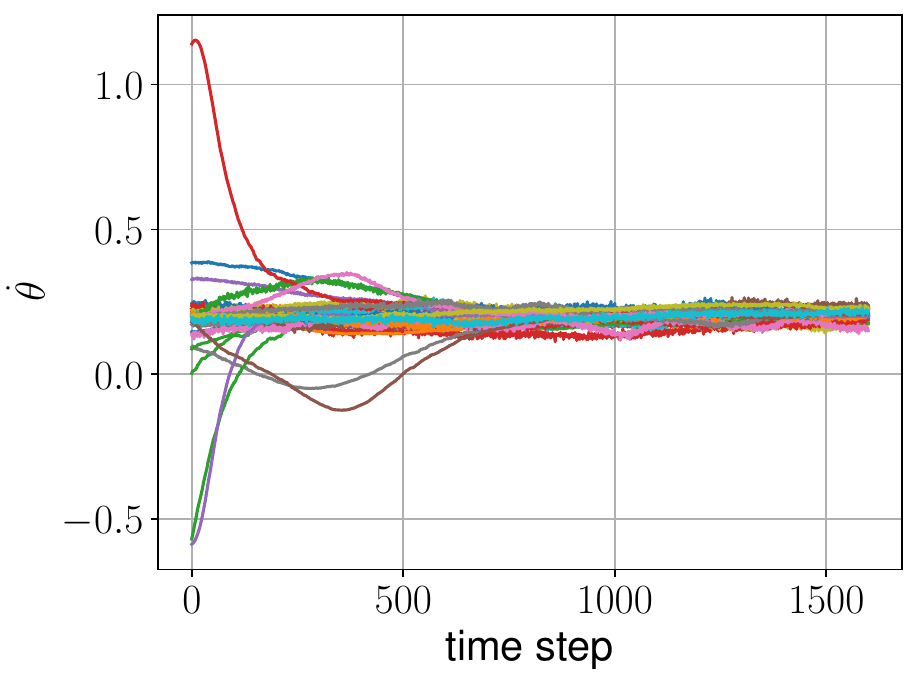}
    \caption{Spectral+ SAC Controller}
    \label{fig:kuramoto_controller_spectral_sac}
  \end{subfigure}
  \caption{Synchronization of frequency ($\dot\theta$) under SAC and Spectral + SAC controller, for 1600 time steps on a single trajectory. Each curve represents a different agent.}
  \label{fig:kuramoto_trajectory}
  \vspace{-10pt}
\end{figure}

%%%%%%%%%%%%%%%%%%%%%%%%%%%%%%%%%%%%%%%%%%%%%%%%%%%%%%%%%%
\section{Conclusion}\label{sec:conclusion}
 \vspace{-0.25em}
%%%%%%%%%%%%%%%%%%%%%%%%%%%%%%%%%%%%%%%%%%%%%%%%%%%%%%%%%%
In this work, utilizing local spectral representations, we provide the first provably efficient algorithm for scalable network control in continuous state-action spaces. We validate our results numerically, where we find that utilizing $\kappa$-local spectral features can achieve effective control on a thermal network control problem as well as a Kuramoto nonlinear coupled oscillator control problem. Moreover, in both cases, we demonstrate that our approach has benefits over generic neural network approximations for local $Q_i$-value functions. Collectively, our theoretical and empirical results demonstrate the validity and importance of a representation-based viewpoint in achieving more effective and scalable control in continuous state-action network MDPs.

\bibliography{references, networked_references, Zotero_references,RL_references}

%%%%%%%%%%%%%%%%%%%%%%%%%%%%%%%%%%%%%%%%%%%%%%%%%%%%%%%%%%
 \vspace{-0.5em}
\section{Appendix}\label{appendix}
 \vspace{-0.25em}
%%%%%%%%%%%%%%%%%%%%%%%%%%%%%%%%%%%%%%%%%%%%%%%%%%%%%%%%%%

\subsection{Example of thermal control in buildings as network MDP}
\label{appendix:more_network_mdp_examples}

\begin{example}[Thermal control in buildings]
The problem of thermal control of multiple zones in a building can also be cast as a network MDP. Consider a multi-zone building with a Heating Ventilation and Air Conditioning (HVAC) system. Each zone is equipped with a sensor that can measure the local temperatures and can adjust the supply air flow rate of its associated HVAC system. For simplicity, we consider a discrete-time linear thermal dynamics model based on \cite{zhang2016decentralized, li2021distributed}, where for any $i \in [n]$,
\[
x_i(t+1) - x_i(t) = \frac{\Delta}{v_i \zeta_i} (\theta^o(t) - x_i(t)) + \sum_{j \in N_i}\frac{\Delta}{v_i \zeta_{ij}} (x_j(t) - x_i(t)) 
+ \frac{\Delta}{v_i} a_i(t) + \frac{\Delta}{v_i} \pi_i + \sqrt{\frac{\Delta}{v_i}} w_i(t),
\]
where $x_i(t)$ denotes the temperature of zone $i$ at time $t$, $a_i(t)$ denotes the control input of zone $i$ that is
related with the air flow rate of the HVAC system, $\theta^o(t)$ denotes the outdoor temperature, $\pi_i$ represents
a constant heat from external sources to zone $i$, $w_i(t)$ represents random disturbances, $\Delta$ is the time
resolution, $v_i$ is the thermal capacitance of zone $i$, $\zeta_i$ represents the thermal resistance of the windows and
walls between zone $i$ and the outside environment, and $\zeta_{ij}$ represents the thermal resistance of the walls
between zone $i$ and $j$. Again, we note that the transition dynamics of each agent depends only on its neighbors (and itself). At each zone $i$, there is a desired temperature $\theta_i^*$ set by the users. The local reward function is composed
of the (negative) deviation from the desired temperature and the control cost, i.e.
\[
r_i(t) = - \left((x_i(t) - \theta_i^*)^2 + \alpha_i a_i(t)^2\right),
\]
where $\alpha_i > 0$ is a trade-off parameter.
    
\end{example}

\subsection{On the exponential decay property}
\label{appendix:exponential_decay}

It may not be immediately clear when the exponential decay property holds. The following lemma (cf. Appendix A in \cite{pmlr-v120-qu20a}) highlights that for a local policy where each agent's actions depend only only on its and its neighbors' states (i.e. $\pi_i(\cdot \mid s) \equiv \pi_i(\cdot \mid s_{N_i})$ ), the exponential decay property holds generally, with \( \rho = \gamma \). We defer the proof to our appendix.

\begin{restatable}{lemma}{expDecayHolds}
\label{lemma:exp_decay_holds}
    Suppose $\forall i \in [n]$, agent $i$ adopts a localized policy, i.e. $\pi_i(\cdot \mid s) \equiv \pi_i(\cdot \mid s_{N_i})$. Suppose also that the local rewards are bounded such that $0 \leq r_i \leq \bar{r}$. Then the \( \left( \frac{\bar{r}}{1 - \gamma}, \gamma \right) \)-exponential decay property holds.
\end{restatable}
We note that under some mixing time assumptions on the MDP~\cite{pmlr-v120-qu20a}, the exponential decay property may in fact hold for \( \rho < \gamma \) depending on the system parameters, making it applicable to problems with large discount factors or even in the average-reward setting~\cite{qu2020scalable}.

We proceed now to prove Lemma \ref{lemma:exp_decay_holds}, which shows that the exponential decay property holds for localized policies and bounded rewards. We note that this was first shown in \cite{pmlr-v120-qu20a}, and we provide the proof here for completeness. 

\begin{proof}
    Consider any $i$, and choose any natural number $\kappa$. For an arbitrary $(s,a) = (s_{N_i^\kappa}, s_{N_{-i}^\kappa}, a_{N_i^\kappa}, a_{N_{-i}^\kappa})$, consider any state-action pair $(s',a')$ that differs with $(s,a)$ only outside the $N_i^\kappa$-neighborhood, i.e. $(s',a') = (s_{N_i^\kappa}, s'_{N_{-i}^\kappa}, a_{N_i^\kappa}, a'_{N_{-i}^\kappa})$. For any natural number $t$, let $p_{t,i}$ denote the distribution of $s_i(t), a_i(t)$ conditional on $s(0) = s, a(0) = a$, and let $p_{t,i}'$ denote the distribution of $s_i(t), a_i(t)$ conditional on $s(0) = s', a(0) = a'$. Then,
    \begin{align*}
        Q_i^\pi(s,a) - Q_i^\pi(s',a') = & \ \xpt{\sum_{t=0}^\infty \gamma ^t r_i(s_i(t),a_i(t)) \mid s(0) = s, a(0) = a} - \xpt{\sum_{t=0}^\infty \gamma ^t r_i(s_i(t),a_i(t)) \mid s(0) = s', a(0) = a'} \\
        \labelrel{=}{eq:used_p_ti_defn} & \ \sum_{t=0}^\infty \gamma^t \left(\xpt[p_{t,i}]{r_i(s_i(t),a_i(t))} - \xpt[p_{t,i}']{r_i(s_i(t),a_i(t))}\right) \\
        = & \ \sum_{t=0}^\kappa \gamma^t \left(\xpt[p_{t,i}]{r_i(s_i(t),a_i(t))} - \xpt[p_{t,i}']{r_i(s_i(t),a_i(t))}\right) \\
        & \quad + \sum_{t=\kappa+1}^\infty \gamma^t \left(\xpt[p_{t,i}]{r_i(s_i(t),a_i(t))} - \xpt[p_{t,i}']{r_i(s_i(t),a_i(t))}\right) \\
        \labelrel{=}{eq:p_p'_first_kappa_same} & \ \sum_{t=\kappa+1}^\infty \gamma^t \left(\xpt[p_{t,i}]{r_i(s_i(t),a_i(t))} - \xpt[p_{t,i}']{r_i(s_i(t),a_i(t))}\right)  \\
        \labelrel{\leq}{eq:use_r_bar} & \ \frac{\gamma^{\kappa+1}}{1-\gamma} \bar{r}.
    \end{align*}
    Above,  (\ref{eq:used_p_ti_defn}) is a direct application of the definition of $p_{t,i}$ and $p_{t,i}'$. Meanwhile, (\ref{eq:p_p'_first_kappa_same}) utilizes the fact that for any $0\leq t \leq \kappa$, we have $p_{t,i} \equiv p_{t,i}'$. This is because of (a) localized policy, such that $a_i(t)$ depends only on $s_{N_i}(t-1)$, and (b) factorized localized dynamics, that $s_{N_i^j}(t)$ depends only only on $s_{N_i^{j+1}}(t)$ for any natural number $j$; hence an iterative argument shows that for any $t$, $p_{t,i}$ and $p_{t,i}'$ both only depend on $s_{N_i^t}(0)$ and $a_{N_i^t}(0)$. Thus, since $(s,a)$ and $(s',a')$ share identical $s_{N_i^\kappa}(0)$ and $a_{N_i^\kappa}(0)$, it follows that $p_{t,i} \equiv p_{t,i}'$ for $t \leq \kappa$. Finally, (\ref{eq:use_r_bar}) uses the fact that bounded reward assumption, i.e. $0 \leq r_i \leq \bar{r}$. The proof then concludes by rerunning the argument on $Q_i^\pi(s',a') - Q_i^\pi(s,a)$.
\end{proof}

Next, we state and prove the following elementary technical result, which bounds any two truncated $Q$(or $V$)-functions with different weights. 
\begin{lemma}
\label{lemma:diff_weights_truncated_val_function_bdd}
Suppose the $(c,\rho)$-exponential decay property holds. Then, for any two different weights $w_i(s_{N_i^\kappa}, a_{N_i^\kappa}; s_{N_{-i}^\kappa}, a_{N_{-i}^\kappa})$ and $w'_i(s_{N_i^\kappa}, a_{N_i^\kappa}; s_{N_{-i}^\kappa}, a_{N_{-i}^\kappa})$ over the space $\mathcal{S}_{N_{-i}^\kappa} \times \mathcal{A}_{N_{-i}^\kappa}$, i.e. 
\begin{align*}
    &\sum_{s_{N_{-i}^\kappa}, a_{N_{-i}^\kappa}} w_i(s_{N_i^\kappa}, a_{N_i^\kappa}; s_{N_{-i}^\kappa}, a_{N_{-i}^\kappa}) = 1, \\
    & \sum_{s_{N_{-i}^\kappa}, a_{N_{-i}^\kappa}} w'_i(s_{N_i^\kappa}, a_{N_i^\kappa}; s_{N_{-i}^\kappa}, a_{N_{-i}^\kappa}) = 1,
\end{align*}
we have 
\begin{align*}
    \abs*{\hat{Q}_i^{\pi}(s_{N_i^\kappa}, a_{N_i^\kappa}) - (\hat{Q}')_i^{\pi}(s_{N_i^\kappa}, a_{N_i^\kappa})} \leq 2c\rho^{\kappa+1}, 
\end{align*}
where 
\begin{align*}
    & \hat{Q}_i^{\pi}(s_{N_i^\kappa}, a_{N_i^\kappa}) = \sum_{s_{N_{-i}^\kappa}, a_{N_{-i}^\kappa}} w_i(s_{N_i^\kappa}, a_{N_i^\kappa}; s_{N_{-i}^\kappa}, a_{N_{-i}^\kappa}) Q_i^{\pi}(s_{N_i^\kappa}, s_{N_{-i}^\kappa}, a_{N_i^\kappa}, a_{N_{-i}^\kappa}), \\
    & (\hat{Q}')_i^{\pi}(s_{N_i^\kappa}, a_{N_i^\kappa}) = \sum_{s_{N_{-i}^\kappa}, a_{N_{-i}^\kappa}} w'_i(s_{N_i^\kappa}, a_{N_i^\kappa}; s_{N_{-i}^\kappa}, a_{N_{-i}^\kappa}) Q_i^{\pi}(s_{N_i^\kappa}, s_{N_{-i}^\kappa}, a_{N_i^\kappa}, a_{N_{-i}^\kappa})
\end{align*}
Similarly, for any two different weights 
$w_i(s_{N_i^\kappa}; s_{N_{-i}^\kappa})$ and $w'_i(s_{N_i^\kappa}; s_{N_{-i}^\kappa})$ over the space $\mathcal{S}_{N_{-i}^\kappa}$, i.e. 
\begin{align*}
    &\sum_{s_{N_{-i}^\kappa}} w_i(s_{N_i^\kappa}; s_{N_{-i}^\kappa}) = 1, \\
    & \sum_{s_{N_{-i}^\kappa}} w'_i(s_{N_i^\kappa}; s_{N_{-i}^\kappa}) = 1,
\end{align*}
we have 
\begin{align*}
    \abs*{\hat{V}_i^{\pi}(s_{N_i^\kappa} - (\hat{V}')_i^{\pi}(s_{N_i^\kappa})} \leq 2c\rho^{\kappa+1}, 
\end{align*}
where 
\begin{align*}
    & \hat{V}_i^{\pi}(s_{N_i^\kappa}) = \sum_{s_{N_{-i}^\kappa}} w_i(s_{N_i^\kappa}; s_{N_{-i}^\kappa}) V_i^{\pi}(s_{N_i^\kappa}, s_{N_{-i}^\kappa}), \\
    & (\hat{V}')_i^{\pi}(s_{N_i^\kappa}) = \sum_{s_{N_{-i}^\kappa}} w'_i(s_{N_i^\kappa}; s_{N_{-i}^\kappa}) V_i^{\pi}(s_{N_i^\kappa}, s_{N_{-i}^\kappa})
\end{align*}

\end{lemma}
\begin{proof}
    Compare both truncated $Q$-functions to a $Q$-function evaluated at any specific state action pair where the states and actions of the agents in $N_i^\kappa$ are $s_{N_i^\kappa}$ and $a_{N_i^\kappa}$ respectively. The desired result then follows by Definition \ref{definition:c_rho_property}. A similar argument works for the $V$-function.
\end{proof}

\subsection{Helper results on factorization of network probability transition}
\label{appendix:helper_factorization_prop1_implies_prop2}

\propOneImpliesPropTwo*

\begin{proof}
To see that, observe that 
\begin{align*}
    & \ \prob{s_{N_i^\kappa}(t+1) \mid s_{N_i^{\kappa+1}}(t), a_{N_i^{\kappa+1}}(t)} =  \prod_{j \in N_i^\kappa} \prob{s_j(t+1) \mid s_{N_j}(t), a_{N_j}(t)} \\
    \labelrel{=}{eq:use_assumption_1} & \ \prod_{j \in N_i^\kappa} \langle \bar{\phi}_{j}(s_{N_j}(t),a_{N_j}(t)), \bar{\mu}_{j}(s_j(t+1))\rangle 
    \labelrel{=}{eq:tensor_defn} \inner{\bigotimes_{j \in N_i^\kappa} \bar{\phi}_{j}(s_{N_j}(t),a_{N_j}(t))}{\bigotimes_{j \in N_i^\kappa} \bar{\mu}_j(s_j(t+1))}.
\end{align*}
Above, (\ref{eq:use_assumption_1}) follows from Property \ref{property:spectral}, while (\ref{eq:tensor_defn}) uses the definition of the inner product of two tensor products. Thus, when Property \ref{property:spectral} holds, Property \ref{property:spectral_kappa} holds by setting 
\begin{align*}
% \label{eq:tensor_product_rep}
    \phi_{i,\kappa}(s_{N_i^{\kappa+1}}(t), a_{N_i^{\kappa+1}}(t)) := \bigotimes_{j \in N_i^\kappa}  \bar{\phi}_j(s_{N_j}(t),a_{N_j}(t)), \ \  \mu_{i,\kappa}(s_{N_i^{\kappa}}(t+1)) := \bigotimes_{j \in N_i^\kappa} \bar{\mu}_j(s_j(t+1)).
\end{align*}
\end{proof}

\subsection{Approximation error of spectral features}

\subsubsection{Approximation error when spectral features exactly factorize $\kappa$-hop transition}
\label{appendix:Qi_approx_error_exact}
We recall and prove this result, which bounds the approximation error of using the truncated spectral features to approximate the local $Q_i$-function, in the case when there is no approximation error in the spectral features in representing the $\kappa$-hop transition.
\QiLinearDecomp*

\begin{proof}
    For notational convenience, we omit the $t$ and $(t+1)$ in the parentheses of the state and action notations, and instead use a superscript $^+$ to denote $(t+1)$, e.g. $s^+$ to denote $s(t+1)$. Observe that
    \begin{align*}
        & \ Q_i^\pi(s_{N_i^{\kappa+1}},a_{N_i^{\kappa+1}}, s_{N_{-i}^{\kappa+1}}, a_{N_{-i}^{\kappa+1}} ) \\
        = & \ r_i(s_i,a_i) + \gamma \int_{s_{N_i^{\kappa}}^+} ds_{N_i^\kappa}^+ \prob{s_{N_i^\kappa}^+ \mid s_{N_i^{\kappa+1}}, a_{N_i^{\kappa+1}}} \left(\int_{s_{N_{-i}^{\kappa}}^+} ds_{N_{-i}^{\kappa}}^+ V_i^\pi(s_{N_{i}^{\kappa}}^+,s_{N_{-i}^{\kappa}}^+) \prob{s_{N_{-i}^{\kappa}}^+ \mid s,a} \right) \\
        \labelrel{=}{barV_defn} & \ r_i(s_i,a_i) + \gamma \int_{s_{N_i^{\kappa}}^+}  ds_{N_i^\kappa}^+ \prob{s_{N_i^\kappa}^+ \mid s_{N_i^{\kappa+1}}, a_{N_i^{\kappa+1}}} \hat{V}_i^\pi(s_{N_i^{\kappa}}^+)  \\
        \labelrel{=}{V_hat_V_bar_diff} & \ r_i(s_i,a_i)  + \gamma \int_{s_{N_i^{\kappa}}^+}  ds_{N_i^\kappa}^+ \prob{s_{N_i^\kappa}^+ \mid s_{N_i^{\kappa+1}}, a_{N_i^{\kappa+1}}} \bar{V}_i^\pi(s_{N_i^{\kappa}}^+) + \gamma  \int_{s_{N_i^{\kappa}}^+}  ds_{N_i^\kappa}^+ \prob{s_{N_i^\kappa}^+ \mid s_{N_i^{\kappa+1}}, a_{N_i^{\kappa+1}}} (\hat{V}_i^\pi(s_{N_i^{\kappa}}^+) - \bar{V}_i^\pi(s_{N_i^{\kappa}}^+)) \\
        \labelrel{=}{eq:use_spectral_kappa} & \ r_i(s_i,a_i) + \gamma \int_{s_{N_i^{\kappa}}^+}  ds_{N_i^\kappa}^+ 
        \left\langle \phi_{i,\kappa}(s_{N_i^{\kappa+1}}, a_{N_i^{\kappa+1}}),  \mu_{i,\kappa}(s_{N_i^\kappa}^+)\bar{V}_i^\pi(s_{N_i^{\kappa}}^+) \right\rangle\\
        &\quad + \gamma \int_{s_{N_i^{\kappa}}^+}  ds_{N_i^\kappa}^+ \prob{s_{N_i^\kappa}^+ \mid s_{N_i^{\kappa+1}}, a_{N_i^{\kappa+1}}}(\hat{V}_i^\pi(s_{N_i^{\kappa}}^+) - \bar{V}_i^\pi(s_{N_i^{\kappa}}^+)) 
    \end{align*}
    In (\ref{barV_defn}) above, we used the notation$$\hat{V}_i^\pi(s_{N_i^{\kappa}}^+) :=  \int_{s_{N_{-i}^{\kappa}}^+} ds_{N_{-i}^{\kappa}}^+ V_i^\pi(s_{N_{i}^{\kappa}}^+,s_{N_{-i}^{\kappa}}^+) \prob{s_{N_{-i}^{\kappa}}^+ \mid s,a},$$
    and in (\ref{V_hat_V_bar_diff}), we recall that we defined 
    \begin{align*}
\bar{V}_i^\pi(s_{N_i^{\kappa}}^+) := \int_{s_{N_{-i}^{\kappa}}^+} \frac{d s_{N_{-i}^{\kappa}}^+}{\mathrm{Vol}(\mathcal{S}_{N_{-i}^\kappa})} V_i^\pi(s_{N_i^{\kappa}}^+, s_{{N_{-i}^\kappa}}^+).
    \end{align*}
    Since the $(c,\rho$)-exponential decay property holds, applying Lemma \ref{lemma:diff_weights_truncated_val_function_bdd}, we have
    $$ \abs*{\hat{V}_i^\pi(s_{N_i^{\kappa}}^+)-\bar{V}_i^\pi(s_{N_i^{\kappa}}^+)} \leq 2c \rho^{\kappa+1}.$$
    Thus our desired result holds by setting 
    \begin{align*}
    \bar{Q}_i^\pi(s_{N_i^{\kappa+1}},a_{N_i^{\kappa+1}}) := r_i(s_i,a_i) +  
        \left\langle \phi_{i,\kappa}(s_{N_i^{\kappa+1}}, a_{N_i^{\kappa+1}}), \gamma \int_{s_{N_i^{\kappa}}^+}  ds_{N_i^\kappa}^+ \mu_{i,\kappa}(s_{N_i^\kappa}^+) \bar{V}_i^\pi(s_{N_i^{\kappa}}^+) \right\rangle
    \end{align*}
\end{proof}

\subsubsection{Results on approximation error of random features}
\label{appendix:random_features_approx_error}

We first state the following result on uniform convergence of random Fourier features, adapted from Claim 1 in \cite{rahimi2007random}.

\begin{lemma}[Uniform convergence of Fourier features]
\label{lemma:rf_uniform_convergence}
Let $\mathcal{M}$ be a compact subset of $\mathbb{R}^d$ with diameter $\text{diam}(\mathcal{M})$. Let $k$ be a positive definite shift-invariant kernel $k(x, y) = k(x - y)$. Define the mapping $z$, where 
\[
z = \sqrt{\frac{2}{D}} 
\begin{bmatrix}
\cos(\omega_1^\top x + b_1) & \cdots & \cos(\omega_D^\top x + b_D)
\end{bmatrix},
\]
where $\omega_1,\dots,\omega_D \in \mathbb{R}^d$ are $D$ iid samples from $p$, where $p$ is the Fourier transform of $k$, i.e. $p(\omega) = \frac{1}{2\pi}\int e^{-j\omega^\top \delta} k(\delta) d\delta$, and $b_1,\dots,b_D$ are $D$ are iid samples from $\mathrm{Unif}(0,2\pi)$. We assume that $k$ is suitably scaled such that $p$ is a probability distribution.

Then, for the mapping $z$ defined above, we have
\[
\Pr\left[\sup_{x, y \in \mathcal{M}} |z(x)^\top z(y) - k(x, y)| \geq \epsilon \right] \leq 2^8 \left( \frac{\sigma_p \, \text{diam}(\mathcal{M})}{\epsilon} \right)^2 \exp\left( -\frac{D \epsilon^2}{4(d+2)} \right),
\]
where $\sigma_p^2 = \mathbb{E}_p[\omega^\top \omega]$ is the second moment of the Fourier transform of $k$.

Further,
\[
\sup_{x, y \in \mathcal{M}} |z(x)^\top z(y) - k(x, y)| \leq \epsilon
\]
with probability  at least $1-\delta$ when
\[
D = \Omega \left( \log\left(\frac{\sigma_p \, \text{diam}(\mathcal{M})^2}{\delta \epsilon}\right) \frac{d}{\epsilon^2} \right).
\]
\end{lemma}

\rfApproxError*

\begin{proof}
Observe that $P\left(s^+_{N_i^\kappa} \mid f_{i,\kappa}(s_{N_i^{\kappa+1},a_{N_i^{\kappa+1}}})\right)$ follows the Gaussian distribution $N(0, \sigma^2 I_{\abs*{N_i^\kappa}S })$. For notational convenience, in this proof, we denote $x := f_{i,\kappa}(s_{N_i^{\kappa+1},a_{N_i^{\kappa+1}}})$, $y := s^+_{N_i^\kappa}.$ In addition, in this proof we denote $d := \abs*{N_i^\kappa}S$. For any $ 0\leq \alpha < 1$, observe that
$$P(y \mid x) = \frac{g_\alpha(x)}{\alpha^d} \exp\left(-\frac{\norm*{(1-\alpha^2)y - x}^2}{2\sigma^2(1-\alpha^2)} \right) p_\alpha(y),$$
where $g_\alpha(x) := \exp(\alpha^2\norm*{x}^2/(2\sigma^2(1-\alpha^2)))$, and $p_\alpha(y) := \frac{\alpha^d}{(2\pi\sigma^2)^{d/2}} \exp(-\norm*{\alpha y}^2/(2\sigma^2))$.\footnote{When $\alpha := 0$, this simplifies to $P(y \mid x) \propto \exp\left(-\frac{\norm*{y-x}^2}{2\sigma^2}\right)$. However, we allow a general $0 \leq \alpha < 1$ because it gives greater flexibility both theoretically and empirically.}

Define $\tilde{g}_\alpha := \max_{i \in [n]}\sup_{x \in f_{i,\kappa}(\gS_{N_i^{\kappa+1}},\gA_{N_i^{\kappa+1}})} \frac{g_\alpha(x)}{\alpha^d}$.\footnote{Here we overload notation to use $f_{i,\kappa}(\gS_{N_i^{\kappa+1}},\gA_{N_i^{\kappa+1}})$ to denote $\{f_{i,\kappa}(s_{N_i^{\kappa+1}},a_{N_i^{\kappa+1}}) \}_{(s_{N_i^{\kappa+1}} ,a_{N_i^{\kappa+1}}) \in \gS_{N_i^{\kappa+1}} \times \gA_{N_i^{\kappa+1}}}$}
Observe now that $k_\alpha(z,z') := \exp(-\frac{\norm*{z-z'}^2}{2\sigma^2(1-\alpha^2)})$ is a positive-definite shift-invariant kernel. Hence, by Lemma \ref{lemma:rf_uniform_convergence}, if $m = \Omega\left(\log\left(\frac{d\sigma^{-2} diam(\gS_{N_i^\kappa})^2}{\delta (\epsilon/\tilde{g}_\alpha)} \right) \frac{d \tilde{g}_\alpha^2}{\ep^2}\right)$ it follows that with probability at least $1 - \delta$, 
\begin{align*}
    \sup_{x,y \in \gS_{N_i^\kappa}} \abs*{\bar{\phi}_{i,\kappa}(x)^\top \bar{\mu}_{i,\kappa}(y) - k_\alpha(x,(1-\alpha^2)y)} \leq \frac{\epsilon_P}{\tilde{g}_\alpha},
\end{align*}
where 
\begin{align*}
    & \ \bar{\phi}_{i,\kappa}(x) = \sqrt{\frac{2}{D}}\left\{\cos\left(\frac{w_\ell^\top x}{\sqrt{1-\alpha^2}} + b_\ell \right)\right\}_{\ell=1}^m, \\
    & \ \bar{\mu}_{i,\kappa}(y) = \sqrt{\frac{2}{D}}\left\{\cos\left(\sqrt{1-\alpha^2}y + b_\ell \right)\right\}_{\ell=1}^m,
\end{align*}
and $\{\omega_\ell\}$'s are drawn iid from $N(0,\sigma^{-2} I_d),$ $\{b_\ell\}$'s are drawn iid from $\mathrm{Unif}(0,2\pi)$. It follows then for any $x \in f_{i,\kappa}(\gS_{N_i^{\kappa+1}},\gA_{N_i^{\kappa+1}})$, we have 
\begin{align*}
    \int_y \left|P(y \mid x) - \hat{\phi}_{i,\kappa}(x)^\top \hat{\mu}_{i,\kappa}(y) \right| dy = & \ \int_y \left|\frac{g_\alpha(x)}{\alpha^d}\right|\left|k_\alpha(x,(1-\alpha^2)y) - \bar{\phi}_{i,\kappa}(x)^\top \bar{\mu}_{i,\kappa}(y) \right| \left|p_\alpha(y)\right| dy \\
    \leq  & \ \tilde{g}_\alpha \frac{\epsilon_P}{\tilde{g}_\alpha} \int_{y} p_\alpha(y) dy  = \ep_P
\end{align*}
where we recall that 
$$ \hat{\phi}_{i,\kappa}(x) := \frac{g_\alpha}{\alpha^d} \bar{\phi}_{i,\kappa}(x), \quad \hat{\mu}_{i,\kappa}(y) := p_\alpha(y) \bar{\mu}_{i,\kappa}(y).$$
The proof then follows by rescaling $\epsilon_P := \epsilon_P/n$, and taking a union bound over all $i \in [n]$.

\end{proof}

\subsection{Algorithm analysis - policy evaluation error}
\label{appendix:policy_eval_error}
% \textcolor{blue}{TODO: define $r_i$, also modify the algorithm to LSTD, $\nu$. Runyu Note: Now we assume solving LSTD exactly, I think that this is okay}

For simplicity, we assume throughout the analysis that we are solving the LSTD step in the policy evaluation exactly, i.e. we take the number of least square solves, $T$, to infinite. Moreover, we drop the $i$ subscript in the notation of $\tilde{\phi}_{i,\kappa}$, and use $\nu$ to denote $\nu_{\pi^{(k)}}$. At round $k$, the algorithm output of the policy parameter $w_i$ of agent $i$ is given by
\begin{align*}
    w_i^{(k)} = (M_i^{(k)})^{-1} H_i^{(k)} r_i
\end{align*}
where 
\begin{align*}
    M_i^{(k)} &= \frac{1}{|D_k|}\sum_{s,a,s',a'\in D_k} \tilde{\phi}_{i,\kappa}(s_{N_i^{\kappa+1}}, a_{N_i^{\kappa+1}})\left(\tilde{\phi}_{i,\kappa}(s_{N_i^{\kappa+1}}, a_{N_i^{\kappa+1}}) - \gamma  \tilde{\phi}_{i,\kappa}(s_{N_i^{\kappa+1}}', a_{N_i^{\kappa+1}}')\right)^\top,\\
    H_i^{(k)} &= \frac{1}{|D_k|}\sum_{s,a,s',a'\in D_k} \tilde{\phi}_{i,\kappa}(s_{N_i^{\kappa+1}}, a_{N_i^{\kappa+1}})\tilde{\phi}_{i,\kappa}(s_{N_i^{\kappa+1}}, a_{N_i^{\kappa+1}})^\top.
\end{align*}
For notational convenience, when the context is clear, we drop the $k$-superscript indicating the current round $k$, and denote $w_i := w_i^{(k)}$, $M_i := M_i^{(k)}$ and $H_i := H_i^{(k)}$.

We define an intermediate variable $\tilde w_i$ as follows:
\begin{align*}
    \tilde w_i = \wM_i^{-1} \wH_i r_i
\end{align*}
where
\begin{align*}
    \wM_i &= \bbE_{s,a\sim\nu} \tilde{\phi}_{i,\kappa}(s_{N_i^{\kappa+1}}, a_{N_i^{\kappa+1}})\left(\tilde{\phi}_{i,\kappa}(s_{N_i^{\kappa+1}}, a_{N_i^{\kappa+1}}) - \gamma \bbE_{s', a'\sim P(\cdot|s,a), \pi(\cdot|s')} \tilde{\phi}_{i,\kappa}(s_{N_i^{\kappa+1}}', a_{N_i^{\kappa+1}}')\right)^\top\\
    \wH_i &= \bbE_{s,a\sim\nu} \tilde{\phi}_{i,\kappa}(s_{N_i^{\kappa+1}}, a_{N_i^{\kappa+1}})\tilde{\phi}_{i,\kappa}(s_{N_i^{\kappa+1}}, a_{N_i^{\kappa+1}})^\top
\end{align*}
and further define
\begin{align*}
    \wQ_i(s_{N_i^{\kappa+1}}, a_{N_i^{\kappa+1}}) = \tilde{\phi}_{i,\kappa}(s_{N_i^{\kappa+1}}, a_{N_i^{\kappa+1}})^\top \tilde w_i
\end{align*}

% The real $Q$-function is $Q_i^{\pi(k)}(s,a) = \phi(s,a)^\top w_i^\star$. From Lemma  \ref{lemma:Q_i_linear_decomp} we have that there exists $$\hat Q_i(s_{N_i^{\kappa+1}}, a_{N_i^{\kappa+1}}) = \tilde{\phi}_{i,\kappa}(s_{N_i^{\kappa+1}}, a_{N_i^{\kappa+1}})^\top \hat w_i$$  such that $|Q_i^{\pi(k)}(s,a) - \hat Q_i(s_{N_i^{\kappa+1}}, a_{N_i^{\kappa+1}})|\le 2c\rho^{\kappa+1}$

The real $Q$-function is $Q_i^{\pi(k)}(s,a)$. From Lemma  \ref{lemma:Q_i_linear_decomp} and Lemma \ref{lemma:approx_error_from_approx_P}, we have that there exists $$\hat Q_i(s_{N_i^{\kappa+1}}, a_{N_i^{\kappa+1}}) = \tilde{\phi}_{i,\kappa}(s_{N_i^{\kappa+1}}, a_{N_i^{\kappa+1}})^\top \hat w_i$$  such that with probability at least $1 - \delta$, for every $ i \in [n]$, 
\begin{align}
\label{eq:nu_deltaSA_bound}
    \xpt[\nu]{|Q_i^{\pi(k)}(s,a) - \hat Q_i(s_{N_i^{\kappa+1}}, a_{N_i^{\kappa+1}})|} \leq  2c\rho^{\kappa+1} + \norm*{\frac{\nu}{\nu^o}}_\infty \frac{\ep_P \gamma \bar{r}}{1-\gamma}
\end{align}

\begin{assump}
\begin{align*}
    \|M_i^{-1}\|\le D
\end{align*}    
\end{assump}
\begin{assump}
\begin{align*}
    \|\tilde{\phi}_{i,\kappa}(s_{N_i^{\kappa+1}}, a_{N_i^{\kappa+1}})\| \le L,~~ \forall~ s_{N_i^{\kappa+1}}, a_{N_i^{\kappa+1}}
\end{align*}
\end{assump}

\begin{lemma}[Bellman Error]\label{lemma:Bellman-error}
On the event that condition (\ref{eq:phi_SVD_error_bound}) in Lemma \ref{lemma:approx_error_from_approx_P} holds, for every $i \in [n]$, we have
    \begin{align*}
        \|\tilde w_i - \hat w_i\| \le 2LD\left(c\rho^{\kappa + 1} + \norm*{\frac{\nu}{\nu^o}}_\infty \frac{\gamma \bar{r} \epsilon_P}{1-\gamma}\right)
    \end{align*}
    \begin{proof}
    From Bellman equation we have that
    \begin{align*}
    Q_i^{\pi(k)}(s,a) &= r_i(s_i,a_i) + \gamma\bbE_{s',a'\sim P,\pi}  Q_i^{\pi(k)}(s',a')\\
    \Longrightarrow~~\hat Q_i(s_{N_i^{\kappa+1}}, a_{N_i^{\kappa+1}}) &= r_i(s_i,a_i) + \gamma \bbE_{s',a'\sim P,\pi} \hat Q_i(s_{N_i^{\kappa+1}}',a_{N_i^{\kappa+1}}') +  \Delta(s,a),
    \end{align*}
    where $\Delta(s,a) = -\left(Q_i^{\pi(k)}(s,a) - \hat Q_i(s_{N_i^{\kappa+1}}, a_{N_i^{\kappa+1}})\right) + \gamma \bbE_{s',a'\sim P,\pi}\left( Q_i^{\pi(k)}(s',a')-\hat Q_i(s_{N_i^{\kappa+1}}',a_{N_i^{\kappa+1}}')\right)$.
    Substituting $\hat Q_i(s_{N_i^{\kappa+1}}, a_{N_i^{\kappa+1}}) = \tilde{\phi}_{i,\kappa}(s_{N_i^{\kappa+1}}, a_{N_i^{\kappa+1}})^\top \hat w_i$ into the above equation we have
    \begin{align*}
        \tilde{\phi}_{i,\kappa}(s_{N_i^{\kappa+1}}, a_{N_i^{\kappa+1}})^\top \hat w_i = \tilde{\phi}_{i,\kappa}(s_{N_i^{\kappa+1}}, a_{N_i^{\kappa+1}})^\top r_i + \gamma \bbE_{s',a'\sim P,\pi}\tilde{\phi}_{i,\kappa}(s_{N_i^{\kappa+1}}', a_{N_i^{\kappa+1}}')^\top \hat w_i + \Delta(s,a).
    \end{align*}
    On both side multiply by $\tilde{\phi}_{i,\kappa}(s_{N_i^{\kappa+1}}, a_{N_i^{\kappa+1}})$ and take expectation over ${s,a\sim\nu}$, we have that
    \begin{align*}
        &\bbE_{s,a\sim\nu} \tilde{\phi}_{i,\kappa}(s_{N_i^{\kappa+1}}, a_{N_i^{\kappa+1}})\tilde{\phi}_{i,\kappa}(s_{N_i^{\kappa+1}}, a_{N_i^{\kappa+1}})^\top \hat w_i \\
        &= \bbE_{s,a\sim\nu} \tilde{\phi}_{i,\kappa}(s_{N_i^{\kappa+1}}, a_{N_i^{\kappa+1}})\tilde{\phi}_{i,\kappa}(s_{N_i^{\kappa+1}}, a_{N_i^{\kappa+1}})^\top r_i \\
        &\quad + \gamma \bbE_{s,a\sim\nu}\tilde{\phi}_{i,\kappa}(s_{N_i^{\kappa+1}}, a_{N_i^{\kappa+1}})\bbE_{s',a'\sim P,\pi}\tilde{\phi}_{i,\kappa}(s_{N_i^{\kappa+1}}', a_{N_i^{\kappa+1}}')^\top \hat w_i + \bbE_{s,a\sim\nu} \tilde{\phi}_{i,\kappa}(s_{N_i^{\kappa+1}}, a_{N_i^{\kappa+1}})\Delta(s,a)\\
        \Longrightarrow~~& \wM_i \hat w_i = \wH_i r_i  + \bbE_{s,a\sim\nu} \tilde{\phi}_{i,\kappa}(s_{N_i^{\kappa+1}}, a_{N_i^{\kappa+1}}')\Delta(s,a)
    \end{align*}
    Further, given that
    \begin{align*}
        \wM_i \tilde w_i = \wH_i r_i ,
    \end{align*}
    on the event that condition (\ref{eq:phi_SVD_error_bound}) in Lemma \ref{lemma:approx_error_from_approx_P} holds.
    \begin{align*}
        \hat w_i - \tilde w_i &= \wM_i^{-1} \bbE_{s,a\sim\nu} \tilde{\phi}_{i,\kappa}(s_{N_i^{\kappa+1}}, a_{N_i^{\kappa+1}}')\Delta(s,a)\\
        \Longrightarrow~~ \|\hat w_i - \tilde w_i\| &\le \|\wM_i^{-1} \| \|\bbE_{s,a\sim\nu} \tilde{\phi}_{i,\kappa}(s_{N_i^{\kappa+1}}, a_{N_i^{\kappa+1}}')\Delta(s,a)\|\\
        &\le 2LD\left(c\rho^{\kappa + 1} + \norm*{\frac{\nu}{\nu^o}}_\infty \frac{\epsilon_P \gamma \bar{r}}{1-\gamma}\right),
    \end{align*}
    where for the final inequality we used (\ref{eq:nu_deltaSA_bound}).
    \end{proof}
\end{lemma}

\begin{lemma}[Statistical Error]\label{lemma:statistical-error} Fix an $i \in [n]$ and $k \in [K]$. For sample size $M_s \ge \log\left(\frac{2(m+1)}{\delta}\right)$, we have that with probability at least $1-2\delta$
    \begin{align*}
        \|w_i^{(k)} - \tilde w_i^{(k)}\|\le O\left(\log\left(\frac{(m+1)}{\delta}\right)\right)\frac{D^2L^4}{\sqrt{M_s}}
    \end{align*}
    \begin{proof}
    Again, for notational simplicity we drop the $k$-superscript. 
        We first bound the differences $\|M_i- \wM_i\|$, $\|H_i-\wH_i\|$. Since  
        \begin{align*}
    M_i &= \frac{1}{|D_k|}\sum_{s,a,s',a'\in D_k} \tilde{\phi}_{i,\kappa}(s_{N_i^{\kappa+1}}, a_{N_i^{\kappa+1}})\left(\tilde{\phi}_{i,\kappa}(s_{N_i^{\kappa+1}}, a_{N_i^{\kappa+1}}) - \gamma  \tilde{\phi}_{i,\kappa}(s_{N_i^{\kappa+1}}', a_{N_i^{\kappa+1}}')\right)^\top,\\
    \wM_i &= \bbE_{s,a\sim\nu} \tilde{\phi}_{i,\kappa}(s_{N_i^{\kappa+1}}, a_{N_i^{\kappa+1}})\left(\tilde{\phi}_{i,\kappa}(s_{N_i^{\kappa+1}}, a_{N_i^{\kappa+1}}) - \gamma \bbE_{s', a'\sim P(\cdot|s,a), \pi(\cdot|s')} \tilde{\phi}_{i,\kappa}(s_{N_i^{\kappa+1}}', a_{N_i^{\kappa+1}}')\right)^\top,
\end{align*}
From the Matrix Bernstein inequality (see Lemma \ref{lemma:matrix_bernstein} in Appendix \ref{appendix:concentration}) we have that when $M_s \ge \log\left(\frac{2(m+1)}{\delta}\right)$ with probability at least $1-\delta$
\begin{align*}
    \|M_i - \wM_i\| \le 8L^2\sqrt{M_s^{-1}\log\left(\frac{2(m+1)}{\delta}\right)}\\
    \|H_i - \wH_i\| \le 8L^2\sqrt{M_s^{-1}\log\left(\frac{2(m+1)}{\delta}\right)}
\end{align*}

Thus with probability $1-2\delta $
\begin{align*}
    \|w_i - \tilde w_i\| &= \|M_i^{-1}H_ir_i - \wM_i^{-1}\wH_i r_i\|\\
    &\le \|M_i^{-1} - \wM_i^{-1}\|\|\wH_ir_i\| + \|M_i^{-1}\|\|H_i - \wH_i\|\|r_i\|\\
    &\le \|M_i^{-1}\wM_i^{-1}\|\|M_i -\wM_i \|\|\wH_i\| + \|M_i^{-1}\|\|H_i - \wH_i\|\\
    &\le O\left(\log\left(\frac{(m+1)}{\delta}\right)\right)\frac{D^2L^4 + L^4}{\sqrt{M_s}} \simeq O\left(\log\left(\frac{(m+1)}{\delta}\right)\right)\frac{D^2L^4}{\sqrt{M_s}},
\end{align*}
which completes the proof.
    \end{proof}
\end{lemma}

Combining the above statement we can get the following Lemma for policy evaluation error, which is a restatement of our result in Lemma \ref{lemma:policy_eval_error_main}.
\begin{lemma}[Policy Evaluation Error, restatement of Lemma \ref{lemma:policy_eval_error_main}]
\label{lemma:policy_eval_error_appendix}
Suppose condition (\ref{eq:phi_SVD_error_bound}) in Lemma \ref{lemma:approx_error_from_approx_P} holds. Suppose the sample size $M_s \ge \log\left(\frac{2(m+1)}{\delta/(Kn)}\right)$. Then, with probability at least $1-2\delta$, for every $i \in [n]$ and $k \in [K]$, the ground truth $Q$ function $Q^{\pi(k)}_i(s,a)$ and the truncated $Q$ function learnt in Algorithm \ref{alg:networked_spectral_control} $\hat{Q}_i(s_{N_i^{\kappa+1}}, a_{N_i^{\kappa+1}})$ satisfies, for any distribution $\bar\nu$ on $\gS \times \gA$,
\begin{align*}
   & \ \xpt[\bar\nu]{| Q^{\pi(k)}_i(s,a) - \hat{Q}_i(s_{N_i^{\kappa+1}}, a_{N_i^{\kappa+1}})|} \\
   \le & \ O\left(c\rho L^2 D\rho^{\kappa+1} +\log\left(\frac{(m+1)}{\delta/(Kn)}\right) \frac{D^2L^5}{\sqrt{M_s}} + L\frac{\ep_P \gamma \bar{r}}{1-\gamma}\left( \norm*{\frac{\bar\nu}{\nu^o}}_\infty + \norm*{\frac{\nu_{\pi^{(k)}}}{\nu^o}}_\infty\right)\right),
\end{align*}
\normalsize
where denoting 
$$\tilde{\varphi}_{i,\kappa} :=  \tilde{\phi}_{i,\kappa}(s_{N_i^{\kappa+1}}, a_{N_i^{\kappa+1}}), \tilde{\varphi}_{i,\kappa}' :=  \tilde{\phi}_{i,\kappa}(s'_{N_i^{\kappa+1}}, a'_{N_i^{\kappa+1}}),$$
\begin{align*}
    & D := \max_{i \in [n], k \in [K]} \norm*{(M_i^{(k)})^{-1}}, \quad L := \max_{i \in [n]} \norm*{\tilde{\varphi}_{i,\kappa}}, \mbox{ where} \\
    & M_i^{(k)} :=  \frac{1}{|D_k|}\sum_{s,a,s',a'\in D_k} \tilde{\varphi}_{i,\kappa} (\tilde{\varphi}_{i,\kappa} - \gamma\tilde{\varphi}'_{i,\kappa})^\top.
\end{align*}
\normalsize
\end{lemma}

\begin{proof}
Suppose the condition (\ref{eq:phi_SVD_error_bound}) in Lemma \ref{lemma:approx_error_from_approx_P} holds. Consider any $i \in [n]$ and $k \in [K]$. 
From Lemma \ref{lemma:Bellman-error} and \ref{lemma:statistical-error} we have that with probability at least $1-2\delta$,
\begin{align*}
    &\quad \xpt[\bar\nu]{| Q^{\pi(k)}_i(s,a) - \tilde{\phi}_{i,\kappa}(s_{N_i^{\kappa+1}}, a_{N_i^{\kappa+1}})^\top w_i^{(k)}|}\\
    & \le \xpt[\bar\nu]{|Q_i^{\pi(k)}(s,a) - \hat Q_i(s_{N_i^{\kappa+1}}, a_{N_i^{\kappa+1}})|} + \xpt[\bar\nu]{|\hat Q_i(s_{N_i^{\kappa+1}}, a_{N_i^{\kappa+1}}) - \tilde{\phi}_{i,\kappa}(s_{N_i^{\kappa+1}}, a_{N_i^{\kappa+1}})^\top w_i^{(k)}|}\\
    &\le \left(2c\rho^{\kappa+1} + \norm*{\frac{\bar\nu}{\nu^o}}_\infty\frac{\ep_P \gamma \bar{r}}{1-\gamma} \right)+ \xpt[\bar\nu]{|\tilde{\phi}_{i,\kappa}(s_{N_i^{\kappa+1}}, a_{N_i^{\kappa+1}})^\top (\hat w_i^{(k)} - w_i^{(k)})|}\\
    &\le \left(2c\rho^{\kappa+1} + \norm*{\frac{\bar\nu}{\nu^o}}_\infty\frac{\ep_P \gamma \bar{r}}{1-\gamma} \right) + L\left(\|\hat w_i^{(k)} - \tilde w_i^{(k)}\| + \|\tilde w_i^{(k)} - w_i^{(k)}\|\right)\\
    &\le \left(2c\rho^{\kappa+1} + \norm*{\frac{\bar\nu}{\nu^o}}_\infty\frac{\ep_P \gamma \bar{r}}{1-\gamma} \right) + L\left(2LD(c\rho^{\kappa + 1} + \norm*{\frac{\nu}{\nu^o}}_\infty\frac{\ep_P \gamma \bar{r}}{1-\gamma}) +O\left(\log\left(\frac{(m+1)}{\delta}\right)\right)\frac{D^2L^4}{\sqrt{M_s}}\right)\\
    & = O\left(c\rho L^2 D\rho^{\kappa+1} +\log\left(\frac{(m+1)}{\delta}\right) \frac{D^2L^5}{\sqrt{M_s}} + L\frac{\ep_P \gamma \bar{r}}{1-\gamma} \left( \norm*{\frac{\bar\nu}{\nu^o}}_\infty + \norm*{\frac{\nu}{\nu^o}}_\infty\right)\right).
\end{align*}
The desired result then follows by rescaling $\delta := \delta/(Kn)$ and taking an union bound over all $i \in [n]$ and $k \in [K]$.
\end{proof}

% \begin{lemma}\label{lemma:auxiliary}
%     Suppose $\{(U_k\in\bbR^d, V_k\in\bbR^d)\}_{k=1}^n$ are i.i.d variables and that 
%     \begin{align*}
%         \|U_kV_k^\top-\bbE U_k V_k^\top\|\le M,
%     \end{align*}
%     then with probability at least $1-\delta$,
%     \begin{align*}
%         \left\|\sum_{k=1}^n \left(U_k V_k^\top - \bbE U_k V_k^\top\right)\right\| \le \sqrt{2\log\left(\frac{2}{\delta}\right)} \frac{M}{\sqrt{n}}.
%     \end{align*}
%     \begin{proof}
%         For any $\alpha \in \bbR^d$ with $\|\alpha\|=1$, we have that
%         \begin{align*}
%             \bbE \left(U_k V_k^\top - \bbE U_k V_k^\top\right)\alpha = 0, ~~ \|\left(U_k V_k^\top - \bbE U_k V_k^\top\right)\alpha\| \le M,
%         \end{align*}
%         then from Azuma concentration inequality with probability at least $1-\delta$
%         \begin{align*}
%             &\sum_{k=1}^n \left(U_k V_k^\top - \bbE U_k V_k^\top\right)\alpha \le \sqrt{2\log\left(\frac{2}{\delta}\right)} \frac{M}{\sqrt{n}},
%         \end{align*}
%         Thus
%         \begin{align*}
%             \left\|\sum_{k=1}^n \left(U_k V_k^\top - \bbE U_k V_k^\top\right)\right\| = \sup_{\|\alpha\|=1} \sum_{k=1}^n \left(U_k V_k^\top - \bbE U_k V_k^\top\right)\alpha \le \sqrt{2\log\left(\frac{2}{\delta}\right)} \frac{M}{\sqrt{n}}
%         \end{align*}
%     \end{proof}
% \end{lemma}
\subsection{Policy gradient analysis}
We show now that our algorithm can find an approximate stationary point of the averaged discounted cumulative reward function $J(\pi^{(\theta)})$. For notational convenience, for a given set of policy parameters $\theta = (\theta_1,\dots,\theta_n)$, we define 
\begin{align*}
    J(\theta) := J(\pi^{\theta}) = \bbE_{s \sim \mu_0}\xpt[a(t) \sim \pi^{(\theta)}(\cdot \mid s(t))]{\sum_{t=0}^\infty \gamma^t r(s(t),a(t)) \mid s(0) = s},
\end{align*}
where we recall that $r(s,a) := \frac{1}{n} \sum_{i=1}^n r_i(s_i,a_i)$.     From Lemma \ref{lemma:policy_gradient}, we have that
    \begin{align*}
        \nabla_{\theta} J(\theta) = & \  \xpt[s \sim d^\theta, a \sim \pi^\theta(\cdot \mid s)]{Q^{\theta}(s,a) \nabla_{\theta_i} \log \pi_i^{\theta}(a_i \mid s_{N_i^{\kappa_\pi}})} \\
        = & \  \xpt[s \sim d^\theta, a \sim \pi^\theta(\cdot \mid s)]{ \frac{1}{n} \sum_{j=1}^n Q_j^{\theta}(s,a) \nabla_{\theta} \log \pi_i^{\theta_i}(a_i \mid s_{N_i^{\kappa_\pi}})} 
    \end{align*}

We first provide the following result, which shows that assuming Lipschitz continuity of the gradient of the objective function $J$ as well as the gradients of $\log\pi^\theta$, there exists the following bound on the following weighted sum of the squared gradient norms.
\begin{lemma}
    \label{lemma:policy_gradient_smoothness_telescope}
    Suppose that $\nabla J(\theta)$ is $L'$-Lipschitz continuous. Suppose that for each $i \in [n]$, $\norm*{\nabla_{\theta_i} \log \pi_i^{\theta_i}(\cdot \mid \cdot)} \leq L_{i,\pi}$. Denote $L_\pi := \sqrt{\sum_{i=1}^n L_{i,\pi}^2}$. Suppose for each round $k \in [K],$ $\theta^{(k+1)} = \theta^{(k)} - \eta \hat{g}^{(k)}$. Then, 
\begin{align*}
    \frac{1}{K}\sum_{k=1}^{K} \norm*{\nabla J(\theta^{(k)})}^2 \leq \frac{\bar{r}/(1-\gamma)}{\eta K} + \frac{1}{K}\sum_{k=1}^{K} \frac{L_\pi\bar{r}}{1-\gamma} \norm*{\nabla J(\theta^{(k)}) - \hat{g}^{(k)}} + \frac{1}{K}\sum_{k=1}^{K}L'\eta\left(\norm*{\hat{g}^{(k)} - \nabla J(\theta^{(k)})}^2 + \left(\frac{L_\pi \bar{r}}{1-\gamma}\right)^2\right).
\end{align*}
\end{lemma}
\begin{proof}
    By the Lipschitz continuity of $\nabla J(\theta)$, we have 
    \begin{align*}
        J(\theta^{(k+1)}) \geq & \  J(\theta^{(k)}) + \eta \inner{\nabla J(\theta^{(k)})}{\hat{g}^{(k)}} - \frac{L'}{2}\norm*{\eta \hat{g}^{(k)}}^2 \\
        = & \ J(\theta^{(k)}) + \eta \norm*{\nabla J(\theta^{(k)})}^2 + \eta \inner{\nabla J(\theta^{(k)})}{\hat{g}^{(k)} - \nabla J(\theta^{(k)})} - \frac{L'\eta^2}{2}\norm*{\hat{g}^{(k)}}^2 
    \end{align*}
    By rearranging and using a telescoping sum, we obtain 
    \begin{align*}
        & \ \eta\sum_{k=1}^{K} \norm*{\nabla J(\theta^{(k)})}^2 \leq \sum_{k=1}^{K}(J(\theta^{(k+1)}) - J(\theta^{(k)})) + \eta \inner{\nabla J(\theta^{(k)})}{\nabla J(\theta^{(k)}) - \hat{g}^{(k)}} + \frac{L'\eta^2}{2}\norm*{\hat{g}^{(k)}}^2 \\
        \implies & \ \frac{1}{K}\sum_{k=1}^{K} \norm*{\nabla J(\theta^{(k)})}^2 \leq \frac{J(\theta^{(K+1)}) - J(\theta^{(1)})}{\eta K} + \frac{1}{K}\sum_{k=1}^{K} \norm*{\nabla J(\theta^{(k)})} \norm*{\nabla J(\theta^{(k)}) - \hat{g}^{(k)}} + \frac{1}{K}\sum_{k=1}^{K}\frac{L'\eta}{2}\norm*{\hat{g}^{(k)}}^2.
    \end{align*}
    Recall that $J(\theta^{(K+1)}) - J(\theta^{(1)}) \leq \bar{r}/(1-\gamma)$.     Hence, by the given assumption on the bound on the derivative term $\nabla_{\theta_i}\log \pi_i^{\theta_i}$, it follows that $\norm*{\nabla J(\cdot)} \leq \frac{L_\pi\bar{r}}{1-\gamma}$. The desired bound then follows by plugging this in as well as using the triangle inequality to decompose $\norm*{\hat{g}^{(k)}}^2$.
\end{proof}

As we can tell from the above result, the crux to bounding the average stationarity gap after $K$ rounds of optimization is the difference between the true gradient $\nabla J(\theta^{(k)})$ and the learned gradient $\hat{g}^{(k)}$ used in the update. In this next result, we bound this error, assuming that the truncated local $\hat{Q}_i$-functions are learned up to some error.

\begin{lemma}
    \label{lemma:learnt_grad_true_grad_err_bound}
For any optimization round $k \in [K]$, let $\hat{\nu}^{(k)}$ denote the empirical distribution of the samples used during round $k$, i.e. $\{s(j),a(j) \}_{j \in [M_s]}$ where $(s(j),a(j)) \sim \nu_{\pi^{(k)}}$. Suppose that for each $\ell \in [n]$, the learnt $\hat{Q}_\ell$-value function satisfies the following error bound: 
    \begin{align}
    \label{eq:xpt_nu_hat_delta_sa_bdd}
        \xpt[\hat\nu^{(k)}]{\abs*{\hat{Q}_\ell^{(k)}(s_{N_\ell^{\kappa+1}},a_{N_\ell^{\kappa+1}}) - Q_\ell(s,a)}} \leq \epsilon_Q
    \end{align}
    Suppose that for each $i \in [n]$, $\norm*{\nabla_{\theta_i} \log \pi_i^{\theta_i}(\cdot \mid \cdot)} \leq L_{i,\pi}$. Denote $L_\pi := \sqrt{\sum_{i=1}^n L_{i,\pi}^2}$. Then, for any optimization round $k \in [K]$, with probability at least $1 - \delta$,
    \begin{align}
        \norm*{\hat{g}^{(k)} - \nabla_{\theta} J(\theta^{(k)})} \leq 2cL_\pi \rho^\kappa + \frac{2\bar{r}L_{\pi}}{1-\gamma}\sqrt{\frac{1}{M_s} \log\left(\frac{d_\theta +1}{\delta/K} \right) } + \epsilon_Q L_\pi,
    \end{align}
    where the $i$-th component of the approximate gradient
    $$\hat{g}_i^{(k)} := \frac{1}{M_s}\sum_{j =1}^{M_s} \frac{1}{n} \sum_{\ell \in N_i^{\kappa +  \kappa_\pi}} \hat{Q}_\ell(s_{N_\ell^{\kappa+1}}(j), a_{N_\ell^{\kappa+1}}(j)) \nabla_{\theta_i} \log \pi_i^{(\theta_i^{(k)})}(a_i(j) \mid s_{N_i^{\kappa_\pi}}(j))$$ 
    is defined in Line \ref{algline:gradient_calc} of Algorithm \ref{alg:networked_spectral_control}. 
\end{lemma}
\begin{proof}
    For notational convenience, in the proof, we fix the optimization round $k \in [K]$, and hence, denote $\hat{g}_i := \hat{g}_i^{(k)}$, $\theta := \theta^{(k)}$ and $\hat{Q}_\ell := \hat{Q}_\ell^{(k)}$ unless otherwise specified. Moreover, we also denote $Q^\theta := Q^{\pi_{\theta}}$ for simplicity.
    From Lemma \ref{lemma:policy_gradient}, for any agent $i \in [n]$, we have that
    \begin{align*}
        \nabla_{\theta_i} J(\theta) = & \  \xpt[s \sim d^\theta, a \sim \pi^\theta(\cdot \mid s)]{Q^{\theta}(s,a) \nabla_{\theta_i} \log \pi_i^{\theta_i}(a_i \mid s_{N_i^{\kappa_\pi}})} \\
        = & \  \xpt[s \sim d^\theta, a \sim \pi^\theta(\cdot \mid s)]{ \frac{1}{n} \sum_{\ell=1}^n Q_\ell^{\theta}(s,a) \nabla_{\theta_i} \log \pi_i^{\theta_i}(a_i \mid s_{N_i^{\kappa_\pi}})} 
    \end{align*}
    To bound the difference between $\hat{g}_i$ and $\nabla_{\theta_i} J(\theta)$, we define the following intermediate terms. 

    We define the terms 
    \begin{align*}
        & \ g_i := \frac{1}{M_s}\sum_{j=1}^{M_s}\left( \frac{1}{n} \sum_{\ell \in N_i^{\kappa + \kappa_\pi}} Q_\ell^{\theta}(s(j),a(j)) \nabla_{\theta_i} \log \pi_i^{\theta_i}(a_i(j) \mid s_{N_i^{\kappa_\pi}}(j))\right) \\
        & \ h_i := \xpt[s \sim d^\theta, a \sim \pi^\theta(\cdot \mid s)]{ \frac{1}{n} \sum_{\ell \in N_i^{\kappa + \kappa_\pi}} Q_\ell^{\theta}(s,a) \nabla_{\theta_i} \log \pi_i^{\theta_i}(a_i \mid s_{N_i^{\kappa_\pi}})}.
    \end{align*}

Then, we decompose the error as 
\begin{align}
\label{eq:gradient_err_decomp}
    \nabla_{\theta_i} J(\theta) - \hat{g}_i = \underbrace{(\nabla_{\theta_i} J(\theta) - h_i)}_{E_{J,h}} + \underbrace{(h_i - g_i)}_{E_{h,g}} + \underbrace{(g_i - \hat{g}_i)}_{E_{g,\hat{g}}}.
\end{align}
We proceed now to bound the three error terms in (\ref{eq:gradient_err_decomp}).

\textbf{Error term $E_{J,h}$.} We can bound the term $E_{J,h}$ as follows. For any $\ell \in [n]$ and positive integer $\kappa$, we define 
    $$\tilde{Q}_\ell^{\theta}(s_{N_\ell^{\kappa}},a_{N_\ell^{\kappa}}) := \sum_{(s_{N_{-\ell}^{\kappa}})',(a_{N_{-\ell}^{\kappa}})'} w((s_{N_{-\ell}^{\kappa}})',(a_{N_{-\ell}^{\kappa}})') Q_\ell(s_{N_\ell^{\kappa}},a_{N_\ell^{\kappa}}, (s_{N_{-\ell}^{\kappa}})',(a_{N_{-\ell}^{\kappa}})'),$$
    where we let $w((s_{N_\ell^{\kappa}})',(a_{N_\ell^{\kappa}})')$ denote the uniform weight over the space $\mathcal{S}_{N_{-\ell}^{\kappa}} \times \mathcal{A}_{N_{-\ell}^{\kappa}} $. From Lemma \ref{lemma:diff_weights_truncated_val_function_bdd}, we know that 
    \begin{align}
    \label{eq:Q_j_minus_tilde_Q_j_err}
        \abs*{Q_\ell^\theta(s,a) - \tilde{Q}_\ell^{\theta}(s_{N_\ell^{\kappa}},a_{N_\ell^{\kappa}})} \leq 2c \rho^{\kappa}.  
    \end{align}
We then have 
    \begin{align*}
        \nabla_{\theta_i} J(\theta) - h_i = & \  \xpt[s \sim d^\theta, a \sim \pi^\theta(\cdot \mid s)]{ \left( \frac{1}{n}\sum_{\ell=1}^n Q_\ell(s,a) - \frac{1}{n} \sum_{\ell \in N_i^{\kappa+\kappa_\pi}} Q_\ell^{\theta}(s,a)  \right)\nabla_{\theta_i} \log \pi_i^{\theta_i}(a_i \mid s_{N_i^{\kappa_\pi}})} \\
        = & \ \xpt[s \sim d^\theta, a \sim \pi^\theta(\cdot \mid s)]{ \left( \frac{1}{n} \sum_{\ell \in N_{-i}^{\kappa+\kappa_\pi}} Q_\ell^{\theta}(s,a)  \right)\nabla_{\theta_i} \log \pi_i^{\theta_i}(a_i \mid s_{N_i^{\kappa_\pi}})} \\
        = & \ \xpt[s \sim d^\theta, a \sim \pi^\theta(\cdot \mid s)]{ \left( \frac{1}{n} \sum_{\ell \in N_{-i}^{\kappa+\kappa_\pi}} \tilde{Q}_\ell^\theta(s_{N_\ell^{\kappa}},a_{N_\ell^{\kappa}}) + \left(Q_\ell^\theta(s,a) - \tilde{Q}_\ell^\theta(s_{N_\ell^{\kappa}},a_{N_\ell^{\kappa}})  \right) \right)\nabla_{\theta_i} \log \pi_i^{\theta_i}(a_i \mid s_{N_i^{\kappa_\pi}})}  \\
        = & \ \xpt[s \sim d^\theta, a \sim \pi^\theta(\cdot \mid s)]{ \left( \frac{1}{n} \sum_{\ell \in N_{-i}^{\kappa+\kappa_\pi}} \left(Q_\ell^\theta(s,a) - \tilde{Q}_\ell^\theta(s_{N_\ell^{\kappa}},a_{N_\ell^{\kappa}})  \right)   \right)\nabla_{\theta_i} \log \pi_i^{\theta_i}(a_i \mid s_{N_i^{\kappa_\pi}})}  \\
         & \quad + \  \xpt[s \sim d^\theta, a \sim \pi^\theta(\cdot \mid s)]{ \left( \frac{1}{n} \sum_{\ell \in N_{-i}^{\kappa+\kappa_\pi}} \tilde{Q}_\ell^\theta(s_{N_\ell^{\kappa}},a_{N_\ell^{\kappa}})  \right)\nabla_{\theta_i} \log \pi_i^{\theta_i}(a_i \mid s_{N_i^{\kappa_\pi}})} \\
         := & \ E_{J,h,1} + E_{J,h,2}
    \end{align*}
    To bound $E_{J,h,1}$, utilizing the bound in (\ref{eq:Q_j_minus_tilde_Q_j_err}) as well as the bound  $\nabla_{\theta_i} \log \pi_i^{\theta_i}(a_i \mid s_{N_i}^\kappa) \leq L_{i,\pi}$ in the statement of the lemma, we have that 
    $$ \norm*{E_{J,h,1}} \leq 2cL_{i,\pi} \rho^{\kappa}.$$
    Meanwhile, observe that by definition, for any $\ell \in N_{-i}^{\kappa+\kappa_\pi}$,  $\tilde{Q}_\ell^\theta(s_{N_\ell^{\kappa}},a_{N_\ell^{\kappa}})$ does not depend on $s_{N_i^{\kappa_\pi}}$. Hence, 
    \begin{align*}
        E_{J,h,2} = & \ \xpt[s \sim d^\theta, a \sim \pi^\theta(\cdot \mid s)]{ \left( \frac{1}{n} \sum_{\ell \in N_{-i}^{\kappa+\kappa_\pi}} \tilde{Q}_\ell^\theta(s_{N_\ell^{\kappa}},a_{N_\ell^{\kappa}})  \right)\nabla_{\theta_i} \log \pi_i^{\theta_i}(a_i \mid s_{N_i^{\kappa_\pi}})} \\
        = & \ \xpt[s \sim d^\theta, a_{-i} \sim \pi^\theta(\cdot \mid s)]{ \left( \frac{1}{n} \sum_{\ell \in N_{-i}^{\kappa+\kappa_\pi}} \tilde{Q}_\ell^\theta(s_{N_\ell^{\kappa}},a_{N_\ell^{\kappa}})  \right) \xpt[a_i \sim \pi_i^{\theta_i}(\cdot \mid s_{N_i^{\kappa_\pi}})]{\nabla_{\theta_i} \log \pi_i^{\theta_i}(a_i \mid s_{N_i^{\kappa_\pi}})}} \\
        = & \ \xpt[s \sim d^\theta, a_{-i} \sim \pi^\theta(\cdot \mid s)]{ \left( \frac{1}{n} \sum_{\ell \in N_{-i}^{\kappa+\kappa_\pi}} \tilde{Q}_\ell^\theta(s_{N_\ell^{\kappa}},a_{N_\ell^{\kappa}})  \right) \nabla_{\theta_i}\left(\int_{a_i} \pi_i^{\theta_i}(a_i \mid s_{N_i^{\kappa_\pi}})da_i\right) } \\
        = & \ \xpt[s \sim d^\theta, a_{-i} \sim \pi^\theta(\cdot \mid s)]{ \left( \frac{1}{n} \sum_{\ell \in N_{-i}^{\kappa+\kappa_\pi}} \tilde{Q}_\ell^\theta(s_{N_\ell^{\kappa}},a_{N_\ell^{\kappa}})  \right) \nabla_{\theta_i} (1) }  = 0.
    \end{align*}
    Thus $E_{J,h,2} = 0$. This implies then that 
    \begin{align}
        \label{eq:E_Jh_bdd}
        \norm*{E_{J,h}} = \norm*{\nabla_{\theta_i}J(\theta) - h_i} \leq 2cL_{i,\pi} \rho^{\kappa}.
    \end{align}

    \textbf{Error term $E_{h,g}$}. To bound $E_{h,g}$, we may use standard concentration inequalities. Observe that 
    \begin{align*}
        E_{h,g} := & \  h_i - g_i \\
        = & \ h_i - \frac{1}{M_s}\sum_{j=1}^{M_s}\left( \frac{1}{n} \sum_{\ell \in N_i^{\kappa + \kappa_\pi}} Q_\ell^{\theta}(s(j),a(j)) \nabla_{\theta_i} \log \pi_i^{\theta_i}(a_i(j) \mid s_{N_i^{\kappa_\pi}}(j))\right) \\
        = & \  \frac{1}{M_s}\sum_{j=1}^{M_s}  \underbrace{\left(h_i- \left( \frac{1}{n} \sum_{\ell \in N_i^{\kappa + \kappa_\pi}} Q_\ell^{\theta}(s(j),a(j)) \nabla_{\theta_i} \log \pi_i^{\theta_i}(a_i(j) \mid s_{N_i^{\kappa_\pi}}(j))\right)\right)}_{E_{h,g}(j)}.
    \end{align*}
    Since 
    \begin{align*}
        h_i = \xpt[s \sim d^\theta, a \sim \pi^\theta(\cdot \mid s)]{ \frac{1}{n} \sum_{\ell \in N_i^{\kappa + \kappa_\pi}} Q_\ell^{\theta}(s,a) \nabla_{\theta_i} \log \pi_i^{\theta_i}(a_i \mid s_{N_i^{\kappa_\pi}})},
    \end{align*}
    it follows that $\xpt{E_{h,g(j)}} = 0$. Moreover, using the fact that for any $\ell \in [n]$, $\theta$ and $(s,a)$ pair, $0 \leq Q_\ell^\theta(s,a) \leq \frac{\bar{r}}{1-\gamma}$, and the bound $\nabla_{\theta_i} \log \pi_i^{\theta_i}(a_i \mid s_{N_i^\kappa}) \leq L_{i,\pi}$ in the assumption, we have $\norm*{E_{h,g}(j)} \leq \frac{2\bar{r}L_\pi}{1-\gamma}$. Using the i.i.d.
    assumption between the samples $j \in [M_s]$, we may apply Bernstein's concentration inequality for vectors (see Lemma \ref{lemma:matrix_bernstein}) to find that for any $\delta > 0$, with probability at least $1 - \delta$, 
    \begin{align}
    \label{eq:E_hg_bdd}
        \norm*{E_{h,g}} \leq \frac{2\bar{r}L_{i,\pi}}{1 - \gamma} \sqrt{ \frac{1}{M_s}\log\left(\frac{d_\theta+1}{\delta}\right)},
    \end{align}
    where $d_\theta$ is the dimension of $\theta_i$.
    
\textbf{Error term $E_{g,\hat{g}}$.} Observe that
\begin{align*}
    g_i - \hat{g}_i = & \ \frac{1}{M_s}\sum_{j=1}^{M_s} \frac{1}{n} \sum_{\ell \in N_i^{\kappa + \kappa_\pi}} \left(Q_\ell(s(j),a(j)) - \hat{Q}_\ell(s_{N_\ell^{\kappa+1}}(j), a_{N_\ell^{\kappa+1}}(j))\right)\nabla_{\theta_i}\log \pi_i^{(\theta_i^{(k)})}(a_i(j) \mid s_{N_i^{\kappa_\pi}}(j)) \\
    = & \ \frac{1}{n} \sum_{\ell \in N_i^{\kappa + \kappa_\pi}} \frac{1}{M_s}\sum_{j=1}^{M_s}  \left(Q_\ell(s(j),a(j)) - \hat{Q}_\ell(s_{N_\ell^{\kappa+1}}(j), a_{N_\ell^{\kappa+1}}(j))\right)\nabla_{\theta_i}\log \pi_i^{(\theta_i^{(k)})}(a_i(j) \mid s_{N_i^{\kappa_\pi}}(j)) \\
    \leq & \ \frac{1}{n} \sum_{\ell \in N_i^{\kappa + \kappa_\pi}} \frac{1}{M_s}\sum_{j=1}^{M_s}  \abs*{\left(Q_\ell(s(j),a(j)) - \hat{Q}_\ell(s_{N_\ell^{\kappa+1}}(j), a_{N_\ell^{\kappa+1}}(j))\right)} \norm*{\nabla_{\theta_i}\log \pi_i^{(\theta_i^{(k)})}(a_i(j) \mid s_{N_i^{\kappa_\pi}}(j))} \\
    \labelrel{\leq}{eq:use_nabla_L_bdd_after_CS} & \  \frac{1}{n}\sum_{\ell \in N_i^{\kappa + \kappa_\pi}} \ep_Q \cdot L_{i,\pi} \leq \ep_Q \cdot L_{i,\pi}.
\end{align*}
Above, (\ref{eq:use_nabla_L_bdd_after_CS}) follows from the bound in (\ref{eq:xpt_nu_hat_delta_sa_bdd}), as well as the bound $\norm*{\nabla_{\theta_i}\log \pi_i^{\theta_i}(\cdot \mid \cdot)} \leq L_{i,\pi}$ in the assumption.

Combining the bounds for $E_{J,h}, E_{h,g}$ and $E_{g,\hat{g}}$, we find that with probability at least $1-\delta$,
\begin{align*}
\norm*{\nabla_\theta J(\theta) - \hat{g}} \leq \sqrt{\sum_{i=1}^n \norm*{\nabla_{\theta_i} J(\theta) - \hat{g}_i}^2} \leq 2cL_\pi \rho^\kappa + \frac{2\bar{r}L_{\pi}}{1-\gamma}\sqrt{\frac{1}{M_s} \log\left(\frac{d_\theta +1}{\delta} \right) } + \epsilon_Q L_\pi
 \end{align*}
 The final result then follows by applying a union bound over $k \in [K]$.
\end{proof}

We are now ready to state our main convergence result.
\begin{theorem}[Restatement of Theorem \ref{theorem:main}]
Suppose the sample size $M_s \ge \log\left(\frac{2d_\kappa}{(\delta/Kn)}\right)$. Suppose with probability at least $1 - \delta$, for all $ i \in [n]$,  the following holds for some features $\hat\phi_{i,\kappa}$ and $\hat\mu_{i,\kappa}$:
\begin{align*}
\xpt[\nu^o]{\int_{s^+_{N_i^{\kappa}}}\!\! \left| P(s^+_{N_i^{\kappa}} \!\!\mid\!\! s_{N_i^{\kappa+1}}, a_{N_i^{\kappa+1}}) \!-\! \hat{\phi}_{i,\kappa}(s_{N_i^{\kappa+1}}, a_{N_i^{\kappa+1}})^\top \hat{\mu}_{i,\kappa}(s^+_{N_i^{\kappa}})\right| ds^+_{N_i^{\kappa}}} \leq \ep_P
\end{align*}
for some $\ep_P > 0$. Then, if $\eta = O(1/\sqrt{K})$, we have that with probability at least $1-4\delta$,  
\begin{align*}
\frac{1}{K}\sum_{k=1}^{K} \norm*{\nabla J(\theta^{(k)})}^2 \leq O\left(\frac{\bar{r}/(1-\gamma)}{ \sqrt{K}} + \frac{L_\pi\bar{r}\ep_J}{1-\gamma} + \frac{L'}{\sqrt{K}}\left(\ep_J^2 + \left(\frac{L_\pi \bar{r}}{1-\gamma}\right)^2\right) \right),
\end{align*}
where 
\begin{align*}
    \ep_J := 2cL_\pi \rho^\kappa + \frac{2\bar{r}L_{\pi}}{1-\gamma}\sqrt{\frac{1}{M_s} \log\left(\frac{d_\theta +1}{\delta/K} \right) } + \epsilon_Q L_\pi,
\end{align*}
and
\begin{align*}
    \ep_Q :=  \max_{k \in \{0,1\dots,K-1\}}O\left(c\rho L^2 D\rho^{\kappa+1} +\log\left(\frac{d_\kappa}{\delta}\right) \frac{D^2L^5}{\sqrt{M_s}} + L\frac{\ep_P \gamma \bar{r}}{1-\gamma} \left( \norm*{\frac{\hat\nu^{(k)}}{\nu^o}}_\infty + \norm*{\frac{\nu_{\pi^{(k)}}}{\nu^o}}_\infty\right)\right).
\end{align*}

% the ground truth $Q$ function $Q^{\pi(t)}_i(s,a)$ and the truncated $Q$ function learnt in Algorithm \ref{alg:networked_spectral_control} $\tilde{\phi}_{i,\kappa}(s_{N_i^{\kappa+1}}, a_{N_i^{\kappa+1}})^\top \theta_i$ satisfies, for any distribution $\bar\nu$ on $\gS \times \gA$,
\end{theorem}
\begin{proof}
Fix a $\delta > 0$. Suppose the condition in (\ref{eq:phi_SVD_error_bound}) holds with probability at least $1 - \delta$ for all $i \in [n]$ for a distribution $\nu^o$ over $\gS \times \gA$. In other words, with probability at least $1 - \delta$, for all $ i \in [n]$, the following holds:
\begin{align*}
\xpt[\nu^o]{\int_{s^+_{N_i^{\kappa}}}\!\! \left| P(s^+_{N_i^{\kappa}} \!\!\mid\!\! s_{N_i^{\kappa+1}}, a_{N_i^{\kappa+1}}) \!-\! \hat{\phi}_{i,\kappa}(s_{N_i^{\kappa+1}}, a_{N_i^{\kappa+1}})^\top \hat{\mu}_{i,\kappa}(s^+_{N_i^{\kappa}})\right| ds^+_{N_i^{\kappa}}} \leq \ep_P
\end{align*}
for some $\ep_P > 0$. Then, by Lemma \ref{lemma:policy_eval_error_main}, it follows that with probability at least $1 - 3\delta$, for every $i \in [n]$ and optimization round $k \in [K]$, we have 
\begin{align*}
\xpt[\hat\nu^{(k)}]{\abs*{\hat{Q}_\ell^{(k)}(s_{N_\ell^{\kappa+1}},a_{N_\ell^{\kappa+1}}) - Q_\ell(s,a)}} \leq \epsilon_Q^{(k)},
\end{align*}
where 
$$\ep_Q^{(k)} :=  O\left(c\rho L^2 D\rho^{\kappa+1} +\log\left(\frac{d_\kappa}{\delta/(Kn)}\right) \frac{D^2L^5}{\sqrt{M_s}} + L\frac{\ep_P \gamma \bar{r}}{1-\gamma} \left( \norm*{\frac{\hat\nu^{(k)}}{\nu^o}}_\infty + \norm*{\frac{\nu_{\pi^{(k)}}}{\nu^o}}_\infty\right)\right).$$
Note that by Lemma \ref{lemma:policy_gradient_smoothness_telescope}, with probability at least $1 - \delta$, for every optimization round $k \in [K]$, we have
\begin{align*}
\norm*{\hat{g}^{(k)} - \nabla_{\theta} J(\theta^{(k)})} \leq 2cL_\pi \rho^\kappa + \frac{2\bar{r}L_{\pi}}{1-\gamma}\sqrt{\frac{1}{M_s} \log\left(\frac{d_\theta +1}{\delta/K} \right) } + \epsilon_Q^{(k)} L_\pi. 
\end{align*}
Thus, by picking $\eta = O(1/\sqrt{K})$, using union bound, with probability at least $1 - 4\delta$, we have
\begin{align*}
\frac{1}{K}\sum_{k=1}^{K} \norm*{\nabla J(\theta^{(k)})}^2 \leq O\left(\frac{\bar{r}/(1-\gamma)}{ \sqrt{K}} + \frac{L_\pi\bar{r}\ep_J}{1-\gamma} + L'\eta\left(\ep_J^2 + \left(\frac{L_\pi \bar{r}}{1-\gamma}\right)^2\right) \right),
\end{align*}
where 
\begin{align*}
    \ep_J := 2cL_\pi \rho^\kappa + \frac{2\bar{r}L_{\pi}}{1-\gamma}\sqrt{\frac{1}{M_s} \log\left(\frac{d_\theta +1}{\delta/K} \right) } + \max_{k \in [K]}\epsilon_Q^{(k)} L_\pi.
\end{align*}

\end{proof}

\subsection{Concentration inequalities}
\label{appendix:concentration}

\begin{lemma}[Matrix Bernstein]\label{lemma:matrix_bernstein}
    Suppose $\{M_k\}_{k=1}^n$ are i.i.d random matrices where $M_k \in \bbR^{d_1\times d_2}$ and that
    \begin{align*}
        \|M_k-\bbE M_k\|\le C,
    \end{align*}
    then for a given $\delta\in(0,1)$ and $n \ge \log\left(\frac{d_1 + d_2}{\delta}\right)$, we have
    \begin{align*}
        \textup{Pr}\left(\frac{1}{n}\left\|\sum_{k=1}^n \left(M_k - \bbE M_k\right)\right\| \ge 
 2C\sqrt{n^{-1}\log\left(\frac{d_1 + d_2}{\delta}\right)}\right) \le \delta,
    \end{align*}
    \begin{proof}
        Let $\epsilon:= 2C\sqrt{n^{-1}\log\left(\frac{d_1 + d_2}{\delta}\right)}$, then since $n\ge \log\left(\frac{d_1+ d_2}{\delta}\right)$, we have $\epsilon \le 2C$.

        Now we can apply the matrix Bernstein inequality (Theorem 6.1.1 in \cite{tropp2015introduction}) and get that
        \begin{align*}
            \textup{Pr}\left(\frac{1}{n}\left\|\sum_{k=1}^n \left(M_k - \bbE M_k\right)\right\| \ge 
 \epsilon\right) \le (d_1 + d_2)\exp\left(\frac{-n^2\epsilon^2/2}{nC^2 + Cn\epsilon /3}\right)\\
 \le (d_1 + d_2) \exp\left(\frac{-n^2\epsilon^2/2}{nC^2 + nC^2}\right)
 =(d_1 + d_2)\exp\left(\frac{n\epsilon^2}{4C^2}\right)\\
        \end{align*}
        Substituting $\epsilon= 2C\sqrt{n^{-1}\log\left(\frac{d_1 + d_2}{\delta}\right)}$ into the right hand side of the equation we get
        \begin{align*}
            \textup{Pr}\left(\frac{1}{n}\left\|\sum_{k=1}^n \left(M_k - \bbE M_k\right)\right\| \ge 
 \epsilon\right) \le \delta,
        \end{align*}
        which completes the proof.
    \end{proof}
\end{lemma}

\subsection{Simulation details}
\label{appendix:simulations}

% All code for this project is available as a zip folder with the supplementary material.

\subsubsection{Thermal control of multi-zone building}

\textbf{Problem setup details.} In the simulations, we consider a discrete-time linear thermal dynamics model adapted from \cite{zhang2016decentralized, li2021distributed}, where for any $i \in [n]$,
\[
x_i(t+1) - x_i(t) = \frac{\Delta}{v_i \zeta_i} (\theta^o(t) - x_i(t)) + \sum_{j \in N_i}\frac{\Delta}{v_i \zeta_{ij}} (x_j(t) - x_i(t)) 
+ \frac{\Delta}{v_i} \alpha_i a_i(t) + \sqrt{\frac{\Delta}{v_i}}\beta_i w_i(t),
\]
where $x_i(t)$ denotes the temperature of zone $i$ at time $t$, $a_i(t)$ denotes the control input of zone $i$ that is
related with the air flow rate of the HVAC system, $\theta^o(t)$ denotes the outdoor temperature, $\pi_i$ represents
a constant heat from external sources to zone $i$, $w_i(t)$ represents random disturbances, $\Delta$ is the time
resolution, $v_i$ is the thermal capacitance of zone $i$, $\zeta_i$ represents the thermal resistance of the windows and
walls between zone $i$ and the outside environment, $\zeta_{ij}$ represents the thermal resistance of the walls between zone $i$ and $j$, and $\alpha_i$ and $\beta_i$ denote scaling factors on the input and noise respectively. The local reward is defined as 
\begin{align*}
    r_i(t) = -\rho_i((x_i(t) - \theta_i^*))^2 + a_i(t)^2,
\end{align*}
where $\theta_i^*$ is the target temperature and $\rho_i$ is a trade-off parameter. 

The parameters in the dynamics and rewards are set as follows. For simplicity, we center the temperatures at 0, and hence set the target $\theta_i^*$ to be 0. We set $\rho_i = 3$. We set the following parameters for the dynamics: $\Delta = 20, v_i = 200, \zeta_{ij} = 1, \zeta_i = \frac{1}{2}, \alpha_i = \frac{1}{7}, \beta_i = \sqrt{\frac{v_i}{\Delta}},\theta^0 = 0.$

We also assume $w_i(t)$ to be drawn iid from $N(0,1)$. We set the discount factor in the problem to be 0.75, and (when collecting data) set the horizon length of each episode to be 20.

\textbf{Connectivity. } In this problem, there are $n = 50$ agents, and the agents have circular connectivity and has two neighbors each, such that agent 1 is connected to agents N and agent 2, agent 2 is connected to agents 1 and 3, so on and so forth. 

\textbf{Experimental details. }We assume knowledge of the dynamics and rewards. For policy truncation parameter $\kappa_\pi = 0,1,2,3$, we use $\kappa = 0,1,2,2$ respectively\footnote{We found in practice that using $\kappa = 3$ for $\kappa_\pi = 3$ performed less well in this specific example.} as the evaluation $\kappa$ parameter. We now explain the simulation setup for our implementation of Algorithm \ref{alg:networked_spectral_control} with random features, as well as the benchmark algorithm using a two-hidden layer NN.

\begin{enumerate}
    \item (Spectral embedding generation step). For Algorithm \ref{alg:networked_spectral_control} with random features, for each agent $i$, we use random feature dimension of $m = 30,50,800,800$ for each of the four experiments ($\kappa_\pi = 0,1,2,3$) to represent the function $T^\pi(s_{N_i^{\kappa+1}},a_{N_i^{\kappa+1}}) := \frac{Q_i^\pi(s_{N_i^{\kappa+1}},a_{N_i^{\kappa+1}}) - r_i(s_i,a_i)}{\gamma}$. For the NN implementation, we used a two-hidden layer NN with 128 neurons to represent the function $T^\pi(s_{N_i^{\kappa+1}},a_{N_i^{\kappa+1}})$.

    \item (Policy evaluation step) We used $M_s = 100,200,500,1000$ episodes respectively for each of the four experiments ($\kappa_\pi = 0,1,2,3$) to perform the policy evaluation. For the random features implementation, we used the least squares method in Algorithm \ref{alg:networked_spectral_control} to compute the new weights for the local value functions. For the NN implementation, we ran batch gradient descent, and used a target network with update rate of $0.005$.

    \item (Policy update step). For both implementations, we normalize the policy gradient, and run gradient descent with $\eta = 0.2$.
\end{enumerate}

\subsubsection{Kuramoto synchronization}

\textbf{Problem setup details.} We recall the setup described earlier in the paper. We consider here a Kuramoto system with $n$ agents, with an underlying graph \( \gG = (\mathcal{N}, \mathcal{E}) \), where \( \mathcal{N} = \{1, \ldots, n\} \) is the set of agents and \( \mathcal{E} \subseteq \mathcal{N} \times \mathcal{N} \) is the set of edges. The state of each agent \( i \) is its phase \( \theta_i \in [-\pi,\pi] \), and the action of each agent is a scalar \( a_i \in \gA_i \subset \bbR\) in a bounded subset of $\bbR$. The dynamics of each agent is influenced only by the states of its neighbors as well as its own action, satisfying the following form in discrete time~\cite{mozafari2012oscillator}:

\begin{align*}
    %\label{eq:kuramoto_dynamics_discrete}
    \theta_i(t\!+\!1) \!=\! \theta_i(t) \!+\! dt \!\underbrace{\left(\!\omega_i(t) \!+\! a_i(t)\!+\!\left(\!\sum_{j \in N_i}\! K_{ij} \sin(\theta_j \!-\! \theta_i)\! \right)\!\!\right)}_{:= \dot\theta_i(t)}\! + \epsilon_i(t).
\end{align*}
Above, $\omega_i$ denotes the natural frequency of agent $i$, $dt$ is the discretization time-step, $K_{ij}$ denotes the coupling strength between agents $i$ and $j$, $a_i(t)$ is the action of agent $i$ at time $t$, and $\epsilon_i(t) \sim N(0,\sigma^2)$ is a noise term faced by agent $i$ at time $t$. We note that this fits into the localized transition considered in network MDPs. For the reward, we consider frequency synchronization to a fixed target $\omega_{\mathrm{target}}$. In this case, the local reward of each agent can be described as $r_i(\theta_{N_i},a_i) =  - \left|\dot\theta_i - \omega_{\mathrm{target}}\right|$.

The parameters in the dynamics and rewards are set as follows. We set the target $\omega_{\mathrm{target}}$ to be 0.2. We set the action space as $[-1,1]$. For agents $i$ and $j$ that are connected, we sample $K_{ij}$ uniformly at random from $[0.2,1.2]$. For the natural frequency $\omega_i$'s, we sample them iid uniformly at random from $[-0.5,0.5]$. For the noise, we sample $\epsilon_i(t) \sim N(0,0.0025^2)$. The time resolution is $dt = 0.01$.

We also assume $w_i(t)$ to be drawn iid from $N(0,1)$. We set the discount factor in the problem to be 0.99, and set the horizon length to be 800 steps. 

\textbf{Connectivity. } In this problem, there are $n = 40$ agents, and the agents have circular connectivity and has two neighbors each, such that agent 1 is connected to agents N and agent 2, agent 2 is connected to agents 1 and 3, so on and so forth. 

\textbf{Experimental details (model-free). }In this case, we do not assume access to the dynamics function. We now explain the simulation setup for our implementation of Algorithm \ref{alg:networked_spectral_control} with spectral features, as well as the benchmark algorithm using a two-hidden layer NN.

\begin{enumerate}

    \item (Policy evaluation step)  For both the spectral feature and NN implementation, the features are the last layer of a two-hidden layer neural network with hidden dimension 256. At each iteration, for each agent $i$,  we draw a batch (of 128 transitions) from the replay buffer and we run 1 step of gradient descent on the least square bellman error, and used a target network with update rate of $0.005$.

    \item (Policy update step). For each agent $i$, the policy is parameterized to be a 3-hidden layer NN which outputs the mean and standard deviation of the agent's action, and the input is $s_{N_i}$, i.e. the states of the neighborhood of agent $i$. We update the policy parameters $\{\theta_i\}_{i=1}^n$ by taking one gradient descent step on the following objective:
    \begin{align*}
        J_{\pi}(\theta) = \mathbb{E}_{s \sim \mathcal{D}} \left[ D_{\mathrm{KL}} \left( \prod_{i=1}^n \pi_{\theta_i}(\cdot_i \mid s_{N_i}) \bigg\| \frac{\exp(\sum_{i=1}^n\hat{Q}_i(s_{N_i^{\kappa+1}}, \{\cdot_i\}_{i=1}^n))}{Z(s)} \right) \right],
    \end{align*}
    where $\mathcal{D}$ is a set of data from the replay buffer, $Z(s)$ is a normalization constant. Above, we assume the temperature parameter $\tau$ to be 1. This objective is identical to the implementation in Soft-Actor-Critic (SAC)~\cite{haarnoja2018soft} but for factored policies, as well as using the learned $\hat{Q}_i$-value functions to approximate the value function.

    \item (Feature step). For the spectral features, for each agent $i$, we add an additional feature step, which seeks to regularize the features such that they approximate the top left eigenfunctions of the probability transition $P(s'_{N_i^\kappa} \mid s_{N_i^{\kappa+1},a_{N_i^{\kappa+1}}})$, by taking a gradient descent step on the following objective to update agent $i$'s feature $\phi(s_{N_i^{\kappa+1}},a_{N_i^{\kappa+1}})$:
        \begin{align}
        \min_{\phi = \{\phi_1,\dots,\phi_L\}} \bbE_{d(s,a)}[\|\phi(s_{N_i^{\kappa+1}},a_{N_i^{\kappa+1}})\|^2] - 2\bbE_{d(s,a), s' \sim P(\cdot \mid s,a)} [\omega(s_{N_i^\kappa}')^\top \phi(s_{N_i^{\kappa+1}},a_{N_i^{\kappa+1}})],\label{eq:randomized_functional_SVD_objective}
    \end{align}
    where in practice we pick $d(s,a)$ to be the a set of samples from the current replay buffer. We note that (\ref{eq:randomized_functional_SVD_objective}) can be seen as a randomized way to compute the singular value decomposition (SVD) of the transition operator $P(s'_{N_i^\kappa} \mid s_{N_i^{\kappa+1},a_{N_i^{\kappa+1}}})$, and we picked it due to its better numerical performance compared to existing spectral decomposition methods in the literature~\cite{ren2022spectral}, in our simulation example. We also note that unlike for the model-based case (with random features), we do not have guarantees on the end-to-end performance of the model-free version of the algorithm. However, we note that the feature step encourages the features to minimize the objective in (\ref{eq:phi_SVD_error_bound}). We leave more detailed analysis of this to future work. We give more details on the derivation of (\ref{eq:randomized_functional_SVD_objective}) in Appendix~\ref{appendix:rsvd}. 
\end{enumerate}

\textbf{Experimental details and results (model-based). }In this case, we do assume knowledge of the dynamics function and that the noise is Gaussian, allowing us to use random features as the spectral features.

We focus on discussing the feature generation step, since this is the only difference with the previous model-free case. For the random features, for each agent $i$, we select the random features according to the procedure in Lemma \ref{lemma:spectral_decomp_gaussian} (with feature dimension being 1024), and in the simulations we set $\alpha = 0$. For the NN implementation, we use a two-hidden layer NN with hidden dimension 256. For the NN implementation, for a fair comparison, we also give it knowledge of the dynamics function $f_{i,\kappa}(s_{N_i^{\kappa+1}}, a_{N_i^{\kappa+1}})$, such that it can use this information when computing the local value functions. We note that for the policy evaluation step, for both random features and NN, we perform gradient descent on the Bellman least square error. 

The results of the learning performance are shown below. We see that the spectral-based method displays comparable performance to the NN-based implementation.

\begin{figure}[h]
  \centering
  \includegraphics[width=0.7\hsize]{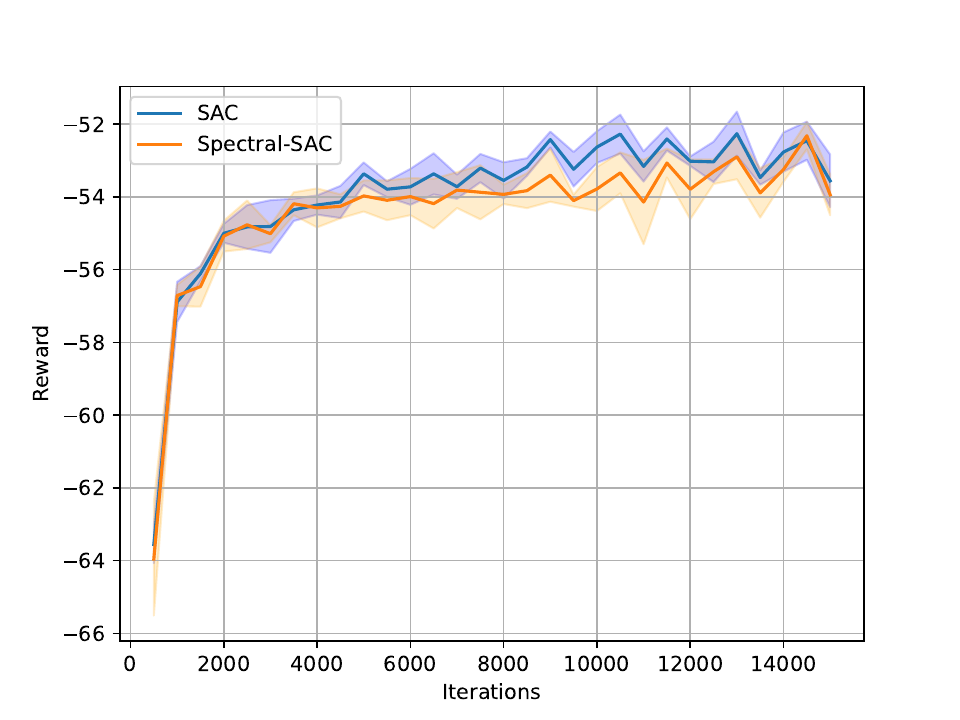}
  \caption{Change in reward during training for Kuramoto oscillator control, $n = 40$, $\kappa_\pi = 1, \kappa = 2$. In this experiment, the dynamics model is known. The performance for each algorithm is averaged over 5 seeds.}
  \label{fig:kuramoto_learning_model_based}
\end{figure}

\subsubsection{Randomized Spectral Decomposition}
\label{appendix:rsvd}
We give here a more detailed derivation of the objective in (\ref{eq:randomized_functional_SVD_objective}). For simplicity, we focus on the single-agent case, when we are trying to decompose $P(s' \mid s,a)$ as $P(s' \mid s,a) \approx \phi(s,a)^\top \mu(s')$, and in particular trying to find the $\phi(s,a)$ in this decomposition. As suggested in \cite{ren2022spectral}, this is akin to finding the top left eigenfunctions of $P(s' \mid s,a)$. Motivated by randomized SVD for computing the top left singular vectors of finite-dimensional matrices, in the functional space setting, we can perform an analogous randomized SVD to learn the top left eigenfunctions of $P(s' \mid s,a)$ according to the following procedure.

\begin{enumerate}
    \item Fix a positive integer $L$.
    \item For each $i \in [L]$, sample a random function $\omega_i(s') \in \bbR$, e.g. $\omega_i(s') = \cos(\alpha_i^\top s' + \beta_i)$, where $\alpha_i \sim N(0,I_S)$ and $\beta_i \sim \mathrm{Unif}([0,2\pi]).$
    \item For each $i \in [L]$, learn a $\phi_i(s,a)$ that approximates $P\omega_i(s,a) := \int_{s'} P(s' \mid s,a) \omega_i(s') ds'$ as follows:
        \begin{enumerate}
            \item Pick a sampling distribution $d(s,a)$, e.g. uniform distribution.
            \item For each $i \in [L]$, solve 
    \begin{align*}
        & \ \min_{\phi_i} \int_{s,a} d(s,a) \left(\phi_i(s,a) - P\omega_i (s,a) \right)^2  \\
        \iff & \  \min_{\phi_i} \int_{s,a} d(s,a) \left(\phi_i(s,a) - \int_{s'}P(s' \mid s,a) \omega_i (s') \right)^2  \\
        \iff &  \ \min_{\phi_i} \int_{s,a} d(s,a) \phi_i(s,a)^2 - 2 \int_{s,a} d(s,a) \int_{s'} ds' P(s' \mid s,a) \omega_i(s') \phi_i(s,a) \\
        \iff & \ \min_{\phi_i} \bbE_{d(s,a)}[\phi_i(s,a)^2] - 2\bbE_{d(s,a), s' \sim P(\cdot \mid s,a)} [\omega_i(s') \phi_i(s,a)]
    \end{align*}
    \end{enumerate}
\end{enumerate}

We note that the final objective is equivalent to solving the $L$ $\phi_i$'s jointly which is single-agent analogue of the objective in (\ref{eq:randomized_functional_SVD_objective}):
\begin{align*}
    \min_{\phi = \{\phi_1,\dots,\phi_L\}} \bbE_{d(s,a)}[\|\phi(s,a)\|^2] - 2\bbE_{d(s,a), s' \sim P(\cdot \mid s,a)} [\omega(s')^\top \phi(s,a)].
\end{align*}

\end{document}